\documentclass[modern]{aastex62}
\usepackage[title]{appendix}
\usepackage{multirow}
\newcommand{\figappendixprefix}{A}
\shorttitle{\textit{NICER} and \textit{Fermi} GBM Observations of Swift~J0243.6+6124}

\begin{document}
\title{\textit{NICER} and \textit{Fermi} GBM Observations of the First Galactic Ultraluminous X-ray Pulsar Swift~J0243.6+6124}
\author{Colleen A. Wilson-Hodge}
\affil{ST12 Astrophysics Branch, NASA Marshall Space Flight Center, Huntsville, AL 35812, USA}
\author{Christian Malacaria}
\affiliation{ST12 Astrophysics Branch, NASA Marshall Space Flight Center, Huntsville, AL 35812, USA}
\affiliation{Universities Space Research Association, NSSTC, Huntsville, AL 35805, USA}\thanks{NASA Postdoctoral Fellow}
\author{Peter A. Jenke}
\affil{University of Alabama in Huntsville, Huntsville, AL 35805, USA}
\author[0000-0002-6789-2723]{Gaurava K. Jaisawal}
\affil{National Space Institute, Technical University of Denmark, Elektrovej, DK-2800 Lyngby, Denmark}
\author{Matthew Kerr}
\affil{Space Science Division, U.S. Naval Research Laboratory, Washington, DC 20375-5352, USA}
\author[0000-0002-4013-5650]{Michael T. Wolff}
\affil{Space Science Division, U.S. Naval Research Laboratory, Washington, DC 20375-5352, USA}
\author{Zaven Arzoumanian}
\affil{Code 662, NASA Goddard Space Flight Center, Greenbelt, MD 20771, USA}
\author[0000-0001-8804-8946]{Deepto Chakrabarty}
\affil{MIT Kavli Institute for Astrophysics and Space Research, Massachusetts Institute of Technology, Cambridge, MA 02139, USA}
\author{John P. Doty}
\affil{Noqsi Aerospace Ltd., Billerica, MA 01821, USA}
\author{Keith C. Gendreau}
\affil{Code 662, NASA Goddard Space Flight Center, Greenbelt, MD 20771, USA}
\author[0000-0002-6449-106X]{Sebastien Guillot} 
\affil{CNRS, IRAP, 9 avenue du Colonel Roche, BP
  44346, F-31028 Toulouse Cedex 4, France} 
\affil{Universit\'e de Toulouse, CNES, UPS-OMP, F-31028 Toulouse, France}
\author{Wynn C. G. Ho}
\affil{Department of Physics and Astronomy, Haverford College, Haverford, PA 19041, USA}
\affil{Mathematical Sciences, Physics and Astronomy, and STAG Research Centre,
University of Southampton, Southampton SO17 1BJ, United Kingdom}
\author{Beverly LaMarr}
\affil{MIT Kavli Institute for Astrophysics and Space Research, Massachusetts Institute of Technology, Cambridge, MA 02139, USA}
\author{Craig B. Markwardt}
\affil{Code 662, NASA Goddard Space Flight Center, Greenbelt, MD 20771, USA}
\author{Feryal \"{O}zel}
\affil{Department of Astronomy, University of Arizona, Tucson, AZ 85721 USA}
\author{Gregory Y. Prigozhin}
\affil{MIT Kavli Institute for Astrophysics and Space Research, Massachusetts Institute of Technology, Cambridge, MA 02139, USA}
\author[0000-0002-5297-5278]{Paul S. Ray}
\affil{Space Science Division, U.S. Naval Research Laboratory, Washington, DC 20375-5352, USA}
\author{Mercedes Ramos-Lerate}
%\affil{Gaia Science Operations Centre (GSOC), European Space Astronomy Centre (ESAC), E-28691, Madrid, Spain}
\affil{Vitrociset Belgium for ESA/ESAC, Camino Bajo del Castillo s/n, 28691 Villanueva de la Cañada, Spain}
\author{Ronald A. Remillard}
\affil{MIT Kavli Institute for Astrophysics and Space Research, Massachusetts Institute of Technology, Cambridge, MA 02139, USA}
\author{Tod E. Strohmayer} 
\affil{Code 662, NASA Goddard Space Flight Center, Greenbelt, MD 20771, USA}
\author{Michael L. Vezie}
\affil{MIT Kavli Institute for Astrophysics and Space Research, Massachusetts Institute of Technology, Cambridge, MA 02139, USA}
\author{Kent S. Wood}
\affil{Praxis Inc., Arlington, VA 22202, USA}
%\author[0000-0001-8804-8946]{Deepto Chakrabarty}
%\author{Beverly LaMarr}
%\author{Gregory Y. Prigozhin}
%\author{Michael L. Vezie}
%\author{Ronald A. Remillard}
%\affil{MIT Kavli Institute for Astrophysics and Space Research, Massachusetts Institute of Technology, Cambridge, MA 02139, USA}
%\author{Craig B. Markwardt}
\collaboration{on behalf of the NICER Science Team}
\noaffiliation

\begin{abstract}
Swift J0243.6+6124 is a newly discovered Galactic Be/X-ray binary, revealed in late September 2017 in a giant outburst with a peak luminosity of $2 \times 10^{39} (d/7\, {\rm kpc})^2$ erg s$^{-1}$ (0.1--10 keV), with no formerly reported activity. At this luminosity, Swift J0243.6+6124 is the first known galactic ultraluminous X-ray pulsar. We describe \textit{Neutron star Interior Composition Explorer (NICER)} and \textit{Fermi} Gamma-ray Burst Monitor (GBM) timing and spectral analyses for this source. A new orbital ephemeris is obtained for the binary system using spin-frequencies measured with GBM and 15--50 keV fluxes measured with the \textit{Neil Gehrels Swift Observatory} Burst Alert Telescope to model the system's intrinsic spin-up.  Power spectra measured with \textit{NICER} show considerable evolution with luminosity, including a quasi-periodic oscillation (QPO) near 50 mHz that is omnipresent at low luminosity and has an evolving central frequency. Pulse profiles measured over the combined 0.2--100 keV range show complex evolution that is both luminosity and energy dependent. Near the critical luminosity of $L\sim 10^{38}$ erg s$^{-1}$, the pulse profiles transition from single-peaked to double peaked, the pulsed fraction reaches a minimum in all energy bands, and the hardness ratios in both \textit{NICER} and GBM show a turn-over to softening as the intensity increases. This behavior repeats as the outburst rises and fades, indicating two distinct accretion regimes. These two regimes are suggestive of the accretion structure on the neutron star surface transitioning from a Coulomb collisional stopping mechanism at lower luminosities to a radiation-dominated stopping mechanism at higher luminosities. This is the highest observed (to date) value of the critical luminosity, suggesting a magnetic field of $B \sim 10^{13}$ G. 
\end{abstract}

\section{Introduction}\label{sec:intro}

Accretion-powered pulsars are X-ray binaries consisting of a highly
magnetized neutron star ($\sim 10^{12}$ G) accreting from an ordinary donor star
\citep{Bildsten_1997}. The long-term evolution of these systems is
largely set by the mass of the donor \citep[see, e.g.,][]{Tauris_vdH_2006}. 
Among high-mass X-ray binaries (HMXBs; donor mass $\gtrsim 8 M_\odot$), 
we may identify two distinct types of systems: persistent or quasi-persistent X-ray sources accreting from
an OB-supergiant donor in a circular or low-eccentricity binary, and
episodic X-ray transients accreting from an Oe/Be-type main sequence
donor in an eccentric binary.  In the latter group are the so-called Be/X-ray
binaries. Two types of outbursts \citep{Stella+86} are generally 
observed from these systems \citep[e.g.,][]{Paul2011, Laplace+17}. Type I or normal outbursts, which generally begin 
around the time of the neutron star's periastron passage, and Type II 
or giant outbursts, which are rarer, longer, reach higher luminosities 
$L\geq 10^{37}$ erg s$^{-1}$, and can start at any orbital phase. 
Until recently the giant outbursts were often considered to be the result 
of an enlarged decretion disk around the Be star, but recent work \citep{Monageng+17} 
shows no correlation between the Be disk radii and the occurrence of 
giant outbursts. However, the normal outbursts were found to
occur generally when the Be disc truncation radius was close to or larger than the 
Roche critical lobe radius at periastron passage.  In this paper, we describe observations of a new Be/X-ray
transient pulsar that is particularly luminous, 
making it a superb target for detailed study.

The X-ray source Swift J0243.6+6124 was initially identified as a new
transient by \textit{Swift} BAT (15--50~keV) on 2017 October 3
\citep{Cenko_2017}, with 0.2--10~keV pulsations with a 9.86 s period also
detected with \textit{Swift} XRT \citep{Kennea_2017}. The onset of the
X-ray outburst was actually detected in the 2--10~keV band a few days
earlier by the \textit{Monitor of All-Sky X-ray Image (MAXI)} sky monitor on September 29, but was
initially misidentified as the nearby source LSI +61$^\circ$ 303
\citep{Sugita_2017a, Sugita_2017b} only 24$^{\prime}$ away. The X-ray pulsations were 
confirmed in data from the \textit{Fermi}
(GBM, 12--25~keV) \citep{Jenke_2017}, \textit{Swift} XRT
\citep{Beardmore_2017} and \textit{NuSTAR}
\citep[3--79~keV;][]{Bahramian_2017,Jaisawal_2018}. \textit{NuSTAR}
observations also showed that the pulse shape is strongly
energy-dependent in the 3--79~keV band, but no cyclotron absorption 
features have yet been detected in the hard X-ray spectrum \citep{Jaisawal_2018}.
Preliminary solutions for a 26--28~d eccentric ($e\approx 0.1$) pulsar
orbit were reported by \citet{Ge_2017} and \citet{Doroshenko_2017}, with an improved
solution given by \citet[][see also Section 4.1 below]{Jenke_2018}.

A variable 6~GHz radio counterpart was also detected,
with a flux of $<$27~$\mu$Jy on October 10 and 76$\pm$7~$\mu$Jy on
November 8 \citep{van_den_Eijnden_2017a, van_den_Eijnden_2017b}. An optical counterpart with magnitude $B=13$, USNO-B1.0 1514+0083050, was identified
by positional coincidence with the {\textit Swift} XRT source
\citep{Kennea_2017}. Optical spectroscopy corresponds to a late
Oe-type or early Be-type star \citep{Kouroubatzakis_2017, Bikmaev+17}. The optical counterpart shows evidence for long-term ($\sim$1000~d) $V$-band variability of order 0.15~mag \citep{Stanek_2017, Kochanek_2017, Nesci_2017}. The distance to the source was intially estimated to be 2.5$\pm$0.5~kpc, based on the estimated absolute magnitude of the stellar companion \citep{Bikmaev+17}.

However, on April 25, 2018, \textit{Gaia} Data Release 2 was made available, including an online catalog\footnote{\url{http://gaia.ari.uni-heidelberg.de/tap.html}} with derived distances for a large number of sources including USNO-B1.0 1514+0083050, with a measured parallax of $0.095 \pm 0.030$ milliarcsec. This catalog was based on \citet{Bailer-Jones2018}, where the \textit{Gaia} team applied best practices to derive distances for all of their stars. USNO-B1.0 1514+0083050 corresponds to source id 465628193526364416. The goal of the \citet{Bailer-Jones2018} study is to provide purely geometric distance estimates, independent of assumptions about the physical properties of or the interstellar extinction towards stars. The distance for Swift J0243.6+6124 from this catalog is 6.8 kpc, with a 1-$\sigma$ range of 5.7--8.4 kpc. A separate analysis, performed by a member of the \textit{Gaia} team, using different priors, resulted in a distance of 8 kpc, with a 5--95\% confidence range of (6.3--12.3 kpc) (M. Ramos, private communication). Since these distances are similar and the $\sim 7$ kpc distance is publicly available, for purposes of this paper, we adopt a distance of 7 kpc based on the \textit{Gaia} measurements.

\begin{figure}
\begin{center}
\includegraphics[width=1.1\textwidth]{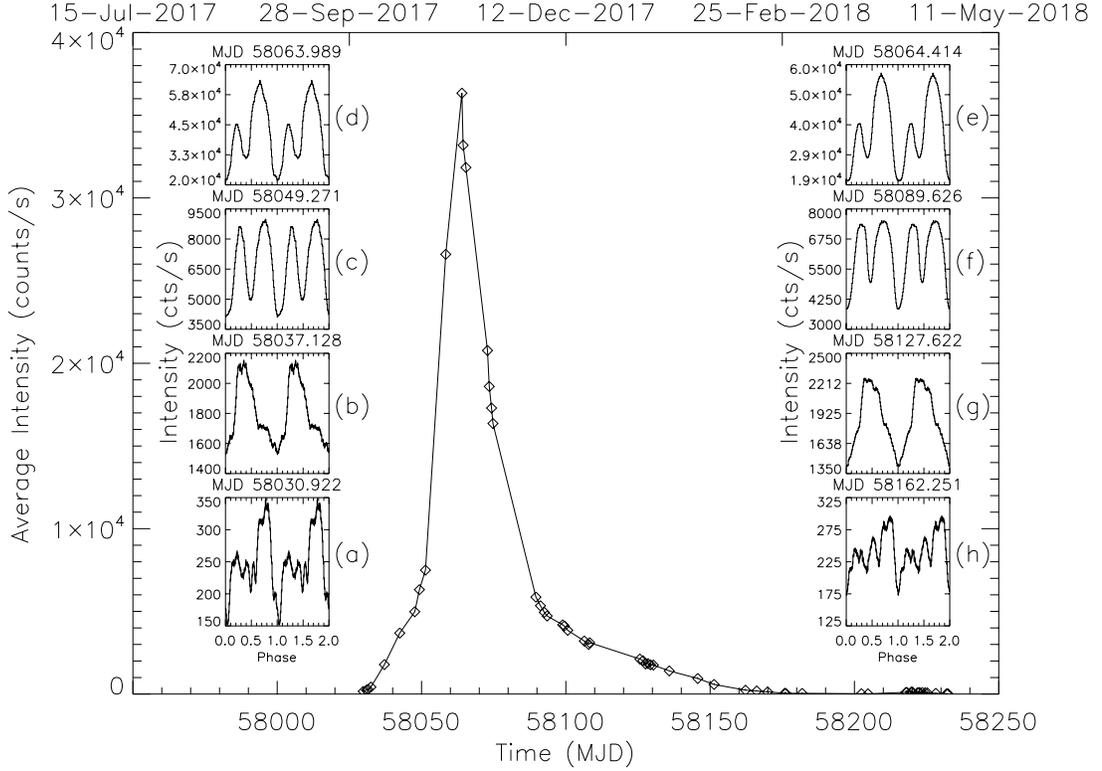}
\end{center}
\caption{Outburst light curve and pulse profile evolution for Swift~J0243.6+6124. The large plot shows the evolution of the average count rate for each \textit{NICER} observation in the 0.2--12 keV range.  The insets to the left and right show the dramatic 0.2--12 keV pulse shape variations as the outburst progresses. For the inset plots, time goes from bottom to top of the page for the rising portion of the outburst (left) and down from top to bottom of the page for the declining part of the outburst (right). }\label{fig:lcandprofiles}
\end{figure}

\begin{figure}
\includegraphics[width=\textwidth]{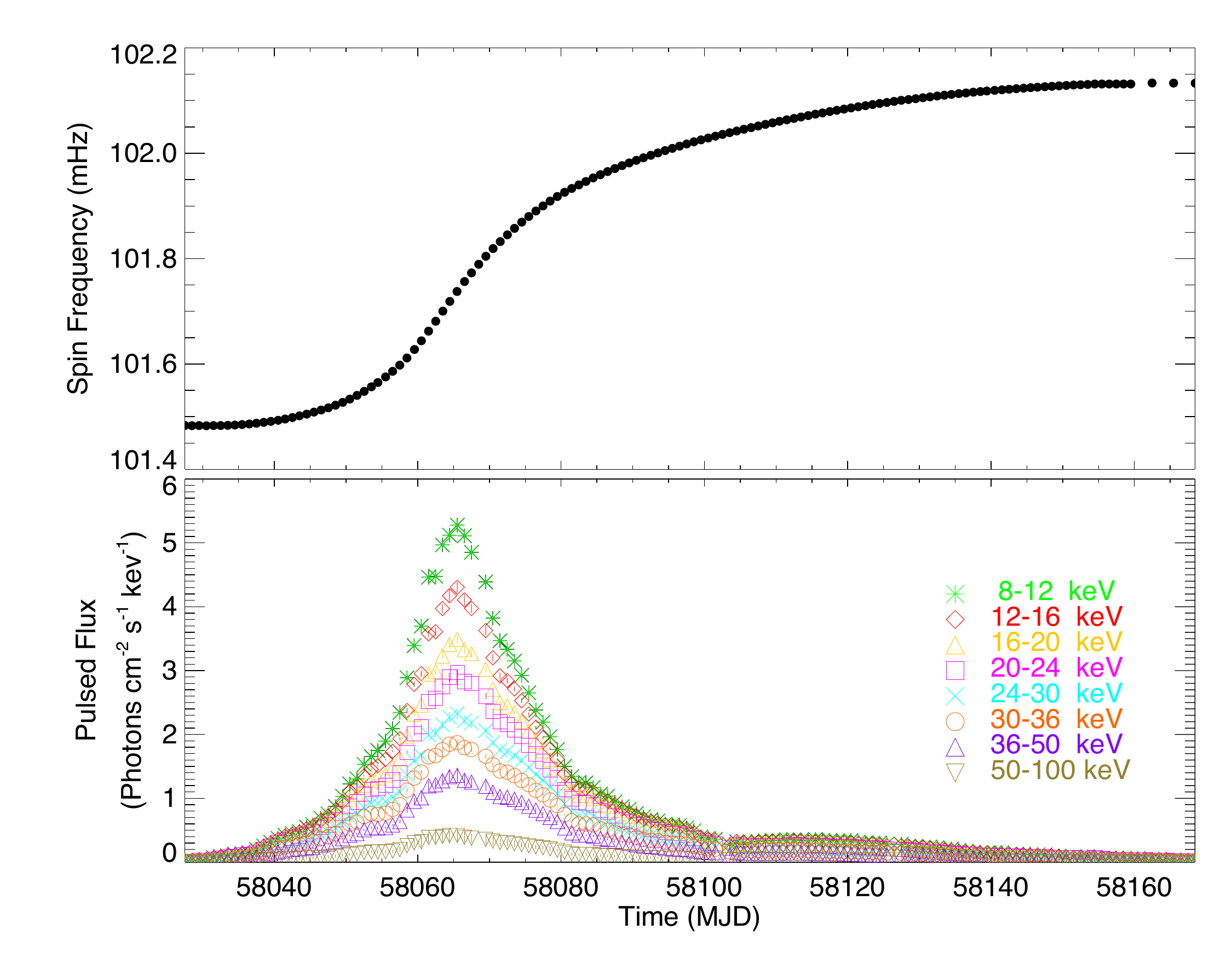}
\caption{Top: Barycentered and orbit-corrected spin-frequency history measured with GBM. Bottom: Pulsed flux measured with GBM \label{gbm:spin_flux} in nine energy bands. Swift J0243.6+6124 was detected from 2017 October 1 through 2018 February 25 (MJD 58027--58174) in single day integrations and through 2018 March 3 (MJD 58180) in 3-day integrations.}
\end{figure}

Transient systems similar to Swift J0243.6+6124 are a particularly valuable laboratory for the study of
magnetically-channeled accretion as they trace through a large dynamic
range of luminosity and mass accretion rate over the course of an
outburst on timescales of days to weeks, before subsiding into X-ray
quiescence. The goal of this paper is to use detailed analyses of the pulse profiles, pulsed fraction, power spectra, and hardness ratios over a broad energy range to understand the changes in accretion column geometry and characteristics as the mass accretion rate evolves in the Swift J0243.6+6124 system.  In this paper, we describe 0.2--12~keV \textit{NICER} and 8-100 keV GBM observations of
Swift J0243.6+6124 obtained throughout the full outburst. Swift J0243.6+6124 evolved over nearly three orders of magnitude in luminosity during this outburst, reaching a peak count rate in the \textit{NICER} energy band of 63,000 cts s$^{-1}$ (see Figure~\ref{fig:lcandprofiles}), nearly 6 times the nominal flux from the Crab nebula and pulsar (11,000 cts s$^{-1}$). As the intensity increased, the pulsar underwent dramatic spin-up, as measured with \textit{Fermi} GBM (see Figure~\ref{gbm:spin_flux}).  We describe the extremely detailed 0.2--12 keV (Section~\ref{subsec:nicerpulse}) and 8--100 keV pulse profile datasets obtained by \textit{NICER} and GBM, respectively, which show considerable evolution with luminosity and energy. Detailed plots of pulse profiles vs. energy as a function of time are shown in the Appendix. Like the pulse profiles, the pulsed fraction (Section~\ref{subsec:nicerpulse}) and power spectra (Section~\ref{subsec:powspec}) measured with \textit{NICER} also evolve with luminosity. Hardness ratio histories measured with \textit{NICER} and GBM (Section~\ref{subsec:hardness}) show a turnover, above which hardness is anti-correlated with luminosity.  Luminosities in the 0.1--10 keV band were estimated from preliminary spectral fits to the \textit{NICER} data (Section~\ref{subsec:spectral}). Implications of these results, which suggest a transition in accretion regimes, are discussed in Section~\ref{sec:discuss}.

\section{Observations of Swift J0243.6+6124}\label{sec:generalobs}

\subsection{\textit{NICER} Instrument and Data}\label{subsec:nicerxti}

The Neutron star Interior Composition Explorer (\textit{NICER}) was launched and installed as an external payload on the International Space Station in 2017 June. \textit{NICER}  contains one instrument, the X-ray Timing Instrument (XTI), sensitive in the 0.2--12 keV range, described in detail in \citet{2016SPIEGendreau} and briefly summarized here. The XTI is composed of an array of 56 X-ray ``concentrator" optics each with an associated focal plane module (FPM) containing a single-pixel detector. Each of the 56 X-ray-concentrator modules consists of 24 nested grazing-incidence gold-coated aluminum foil mirrors, parabolically shaped with a common focal length. Details of the XTI detector subsystem are described in detail in \citet{2016SPIEPrigozhin} and summarized here. The FPMs each consist of a silicon drift detector integrated with a custom pre-amplifier. The FPMs, 52 of which are operating on-orbit, are arranged into 7 groups of 8. Each group of FPMs is controlled by a set of electronics called a Measurement/Power Unit (MPU) slice. Each MPU slice provides the readout and supporting circuitry for its group of 8 FPMs and is completely independent of the other six slices.  

\subsection{\textit{Fermi} GBM Instrument and Data}\label{subsec:GBM_instrument}

The GBM instrument comprises 14 detectors, 12 Sodium Iodide (NaI) detectors sensitive from 8 keV to 1 MeV pointed at various directions in order to cover the whole unocculted sky and two Bismuth germanate (BGO) detectors sensitive from 200 keV to 40 MeV on either side of the  \textit{Fermi} spacecraft \citep{Meegan2009}. Only the NaI detectors are used for this analysis. GBM data are available continuously, when the spacecraft is outside of the South Atlantic Anomaly. There are three continuous data types: CTIME (0.256 s, 8 energy channels) used for the public pulse frequency and pulsed flux measurements (shown in Figure~\ref{gbm:spin_flux} with updates found at \url{https://gammaray.nsstc.nasa.gov/gbm/science/pulsars.html}) and total flux measurements (\url{https://gammaray.nsstc.nasa.gov/gbm/science/earth_occ.html}), CSPEC (4.096 s, 128 energy channels), and CTTE (10 $\mu$s, 128 channels) event by event data, used for this analysis.

\subsection{\textit{Swift} BAT Instrument and Data}\label{subsec:BAT_instrument}

The \textit{Swift} Burst Alert Telescope (BAT) is a hard X-ray monitor \citep{H_Krimm_2013} composed of an array of CdZnTe detectors that are sensitive in the 15--150 keV range with a total detecting area of 5200 cm$^{2}$. It is a coded mask instrument with a 1.4 steradian field-of-view. We use the BAT transient monitor results\footnote{\url{http://heasarc.gsfc.nasa.gov/docs/swift/results/transients/}}  (15--50 keV), provided by the BAT team, to model the accretion torques applied to the neutron star.

\section{\textit{NICER} Observation Analysis}\label{sec:nicerobs}

\subsection{\textit{NICER} Event Deadtime Analysis}\label{subsec:nicerdeadtime}
\begin{deluxetable}{ccccccc}
\tabletypesize{\tiny}	
\tablecolumns{8} 
\tablewidth{0pt}
\tablecaption{\textit{NICER} Observations of Swift~J0243.6+6124 \label{table:nicerobs}} 
\tablehead{
\colhead{NICER ObsID} & \colhead{Date} & \colhead{MJD epoch} & \colhead{On-source} & \colhead{Deadtime corrected} & \colhead{Average rate} & \colhead{Luminosity\tablenotemark{a}} \\ 
      &  & & time (s)& average exposure &0.2--12 keV & 0.1--10 keV  \\
      &  & &         & per MPU slice (s)& (counts s$^{-1}$) & (erg s$^{-1}$) }
\startdata
1050390101 & 2017-10-03 & 58029.805 & 3099 & 2643.4 & 164.8 &  1.19E+37\\
1050390102 & 2017-10-04 & 58030.922 & 4171 & 2782.6 & 239.9 & 1.75E+37 \\
1050390103 & 2017-10-04 & 58031.434 & 21590 & 16645.4 & 281.9 & 2.05E+37\\
1050390104 & 2017-10-06 & 58032.456 & 24029 & 19320.5 & 423.7 & 2.99E+37\\
1050390105 & 2017-10-11 & 58037.128 &2497 & 2198.3 & 1782.9 &  1.08E+38 \\
1050390106 & 2017-10-12 & \nodata & 41 & 0 & \nodata & \nodata \\ 
1050390107 & 2017-10-16 & 58042.455 & 1495 & 420.9 & 3686.8 & 2.01E+38 \\
1050390108 & 2017-10-21 & 58047.663 & 1394 & 615.7 & 4979.0 & 2.86E+38 \\
1050390109 & 2017-10-22 & \nodata & 2334 & 0 & \nodata & \nodata \\
1050390110 & 2017-10-23 & 58049.271 & 900 & 735.4  & 6312.7 &  3.31E+38\\
1050390111 & 2017-10-24 & \nodata & 689 & 0 & \nodata & \nodata \\
1050390112 & 2017-10-25 & 58051.364 & 1003 & 495.8 & 7484.5 &  4.01E+38 \\
1050390113 & 2017-11-02 & 58058.386 & 4566 & 3126.0 & 26583.5 &  1.31E+39\\
1050390114 & 2017-11-06 & 58063.989  & 311 & 257.4 & 36333.8 &  1.77E+39\\
1050390115 & 2017-11-07 & 58064.414 & 1495 & 1276.5 & 33175.4 &  1.66E+39 \\
1050390116 & 2017-11-08 & 58065.381 &3816 & 1915.8 & 31833.4 &  1.58E+39\\
1050390117 & 2017-11-14 &  \nodata & 18 & 0 & \nodata & \nodata \\
1050390118 & 2017-11-15 & 58072.846 & 405 & 374.9 & 20774.0 &  1.32E+39 \\
1050390119 & 2017-11-16 & 58073.460 & 3917 & 2171.9 & 18636.9 &  9.98E+38 \\
1050390120 & 2017-11-17 & 58074.263 & 3916 & 2240.9 & 17297.2 &  1.03E+39 \\
1050390121 & 2017-11-17 & 58074.737 & 1610 & 1517.3 & 16359.3 &  9.40E+38\\
1050390122 & 2017-12-02 & 58089.626 & 3756 & 3676.9 & 5864.4 & 3.25E+38\\
1050390123 & 2017-12-04 & 58091.203 & 1291 & 1016.0 & 5341.7 & 2.70E+38\\
1050390124 & 2017-12-05 & 58092.558 & 1704 & 1590.0 & 4943.2 & 2.71E+38\\
1050390125 & 2017-12-06 & 58093.528 & 1887 & 1854.5 & 4721.4 &  2.65E+38\\
1050390126 & 2017-12-11 & 58098.909 & 210 & 109.0 & 4187.3 &  2.22E+38 \\
1050390127 & 2017-12-12 & 58099.463 & 733 & 640.9 & 4116.8 &  2.23E+38\\
1050390128 & 2017-12-13 & 58100.646 & 482 & 263.5 & 3853.4 &  2.21E+38\\
1050390129 & 2017-12-19 & 58106.402 & 1034 & 594.0 & 3209.5 &  1.84E+38\\
1050390130 & 2017-12-20 & 58107.913 & 528 & 279.7 & 3000.0 & 1.61E+38\\
1050390131 & 2017-12-21 & 58108.237 & 718 & 618.5 & 3110.9 & 1.77E+38\\
1050390132 & 2018-01-07 & 58125.646 & 210 & 208.1 & 2120.3 & 1.29E+38\\
1050390133 & 2018-01-08 & 58126.691 & 990 & 645.3 & 2020.6 &  1.30E+38\\
1050390134 & 2018-01-09 & 58127.622 & 5319 & 3193.7 & 1823.4 &  1.09E+38\\
1050390135 & 2018-01-10 & 58128.492 & 3133 & 2029.3 & 1844.4 &  1.11E+38 \\
1050390136 & 2018-01-10 & 58129.326 & 3288 & 2686.1 & 1775.0 &  1.00E+38\\
1050390137 & 2018-01-12 & 58130.323 & 4541 & 3368.9 & 1755.2 &  1.04E+38\\
1050390138 & 2018-01-17 & 58135.850 & 3230 & 3207.7 & 1399.6 & 8.76E+37\\
1050390139 & 2018-01-27 & 58145.715 & 2266 & 2127.0 & 941.5 &  5.89E+37\\
1050390140 & 2018-02-02 & 58151.352 & 2114 & 2104.8 & 581.6 & 3.83E+37\\
1050390141 & 2018-02-13 & 58162.251  & 3636 & 3492.7 & 236.6 &  1.66E+37\\
1050390142 & 2018-02-17	& 58166.141 & 2768 & 2755.4 & 190.8 &  1.30E+37\\
1050390143 & 2018-02-20 & 58169.900 & 2326 & 1564.1 &  131.5 &  8.57E+36 \\ 
1050390144 & 2018-02-26 & 58175.852 & 2175 & 1805.5 &  49.8 & 3.00E+36\\
1050390145 & 2018-02-27 & 58176.206 & 7058 & 6964.3 &   45.2 &  2.81E+36\\
1050390146 &  2018-03-04 &  58181.9 &  1535 & 1442.7 &  45.7 &  2.49E+36\\
 1050390147 &  2018-03-25 &  58202.4 &  1251 &  1077.7 &  22.0 &   1.20E+36 \\
 1050390148 &  2018-03-27 &  58204.7 &  1705 &  1649.0 &  14.8 &  6.85E+35 \\
 1050390149 & 2018-04-09 &  58217.8 &  5003 & 3966.1 &  104.3 &  6.16E+36 \\
 1050390150 &  2018-04-10 &  58218.1 &   2025 &  1917.4 & 108.7 &  6.22E+36 \\
 1050390151 &  2018-04-11 &  58219.7 &  2278 &  1699.0 &  122.5 &  7.03E+36 \\
 1050390152 &  2018-04-12 &  58220.1 &   1606 &   1111.4 & 120.2 &  6.76E+36 \\
 1050390153 &  2018-04-13 &  58222.0 &  500 & 370.4 &  98.0 &  5.06E+36 \\
 1050390154 &  2018-04-14 &  58222.5 &   19208 &  16463.2 &  102.4 &  5.58E+36 \\
1050390155 &  2018-04-15 &  58223.5 &  14521 &   12004.6 &  90.7 &  4.81E+36 \\
 1050390156 &  2018-04-16 &  58224.5&   7735 &  6281.5 &  78.6 & 4.10E+36 \\
 1050390157 &  2018-04-17 &  58225.3 &  5955 &  5513.6 &  69.8 & 3.57E+36 \\
 1050390158 &  2018-04-20 & 58228.2 &  2785 &   2621.7 &  47.5 &  2.47E+36 \\
1050390159 &  2018-04-23 &  58232.0  &  492 &  459.3 &  24.4 &  9.17E+35 \\
 1050390160 &  2018-04-24 &  58343.4 &   643 &  149.1 & 18.6 & 6.28E+35 \\
\enddata
\tablenotetext{a}{Luminosity calculation assumes isotropic emission and distance = 7 kpc}
\end{deluxetable}

The data analysis procedure for this bright source differs from that used for fainter sources in order to correctly account for detector deadtime. Analysis was done beginning with the unfiltered event files for each MPU slice and the individual slice datasets were kept separate. HEASOFT version 6.22.1 and NICERDAS version 2018-02-22\_V002d were used for this analysis. Good time intervals (GTI) were selected using the {\tt FTOOL} {\tt nimaketime} with the following screening criteria: ISS outside the \textit{NICER}-specific South Atlantic Anomaly (SAA) boundary, \textit{NICER} in tracking mode with pointing direction $<0.015$\degr\ from the source direction with at least 38 detectors enabled, source elevation $>30$\degr\ above the Earth limb, and source direction at least 40\degr\ from the bright Earth. These GTIs were applied to the unfiltered and uncalibrated events for each MPU slice separately using \texttt{niextract-events}. %Note that events within the so-called trumpet filter and within the polar horn regions were not excluded because these backgrounds are negligible for this bright source.

%% To allow the total dead time to be accumulated for each phase bin in the pulse profile for each MPU, the unfiltered, but good time
%% selected files are barycentered using \textit{barycorr} with the DE430 JPL ephemeris. It is important to note that there are 
Two types of deadtime must be separately accounted for in our analysis. First, event-by-event deadtime occurs because processing each event in each detector takes a certain amount of time and that detector is inactive, hence ``dead" while processing an individual event. Second, deadtime can occur because the buffer in each MPU slice reaches its maximum and events are dropped because there is no further room to record them. The first type of deadtime is tracked as a {\tt DEADTIME} column in the \textit{NICER} event files and the second (called GTI exposure in this paper) is tracked in the GTI extensions of the \textit{NICER} event files. Below we account for both types of deadtime in the generation of our \textit{NICER} pulse profiles and light curves.  

The standard \textit{NICER} calibration was applied to the events from each MPU slice separately, using {\tt nimpucal}. This tool produces a Pulse Invariant (PI) column by applying the gain correction specified in the calibration database, \texttt{nixtiflightpi20170601v001.fits}, to the raw pulse height measurements. The gain scale is set so that PI$ = E/10$ eV, where $E$ is the nominal photon energy. Both the calibrated and uncalibrated files are barycentered using {\tt barycorr} with the DE430 JPL ephemeris. The {\tt nimpucal} tool automatically removes events that cannot be calibrated. Because deadtime arises from all events, processing from this point proceeds in parallel for each MPU slice, with the calibrated files for each MPU slice being used to accumulate X-ray events and GTI exposure in the pulse profiles while the uncalibrated files are used to accumulate deadtime. 

Pulse profiles were generated for each observation using the \textit{Fermi} GBM orbital ephemeris described in Section~\ref{subsec:gbmorbit} along with the spin frequency and frequency derivative measured with GBM on the same day as the \textit{NICER} observation, and corrected to the mid-point time of the \textit{NICER} observation. This process was done in a series of steps described in the following paragraphs.

First, source events were selected to form the pulse profiles. Starting with the barycentered, calibrated, but unfiltered event files for each MPU slice, events were selected that were in the chosen energy range (e.g., PI=20--1200 for 0.2--12 keV for the full profiles, PI corresponding to 0.2--1, 1--2, 2--3, 3--5, 5--8, and 8--12 keV for the individual energy bands, and PI corresponding to 4--7 and 7--10 keV for the hardness ratio comparisons with \citealt{Reig+13}). Forced triggers, overshoot, and undershoot events (those events with bits 5, 6, or 7, respectively, set in the {\tt EVENT\_FLAGS} column) were excluded. After correcting the event times for the pulsar's orbit, the pulse phase was computed and events were accumulated in each of 100 phase bins across the pulse period. These folded event profiles were accumulated separately for each MPU slice. 

\begin{figure}
	\begin{center}
     \includegraphics[width=4in]{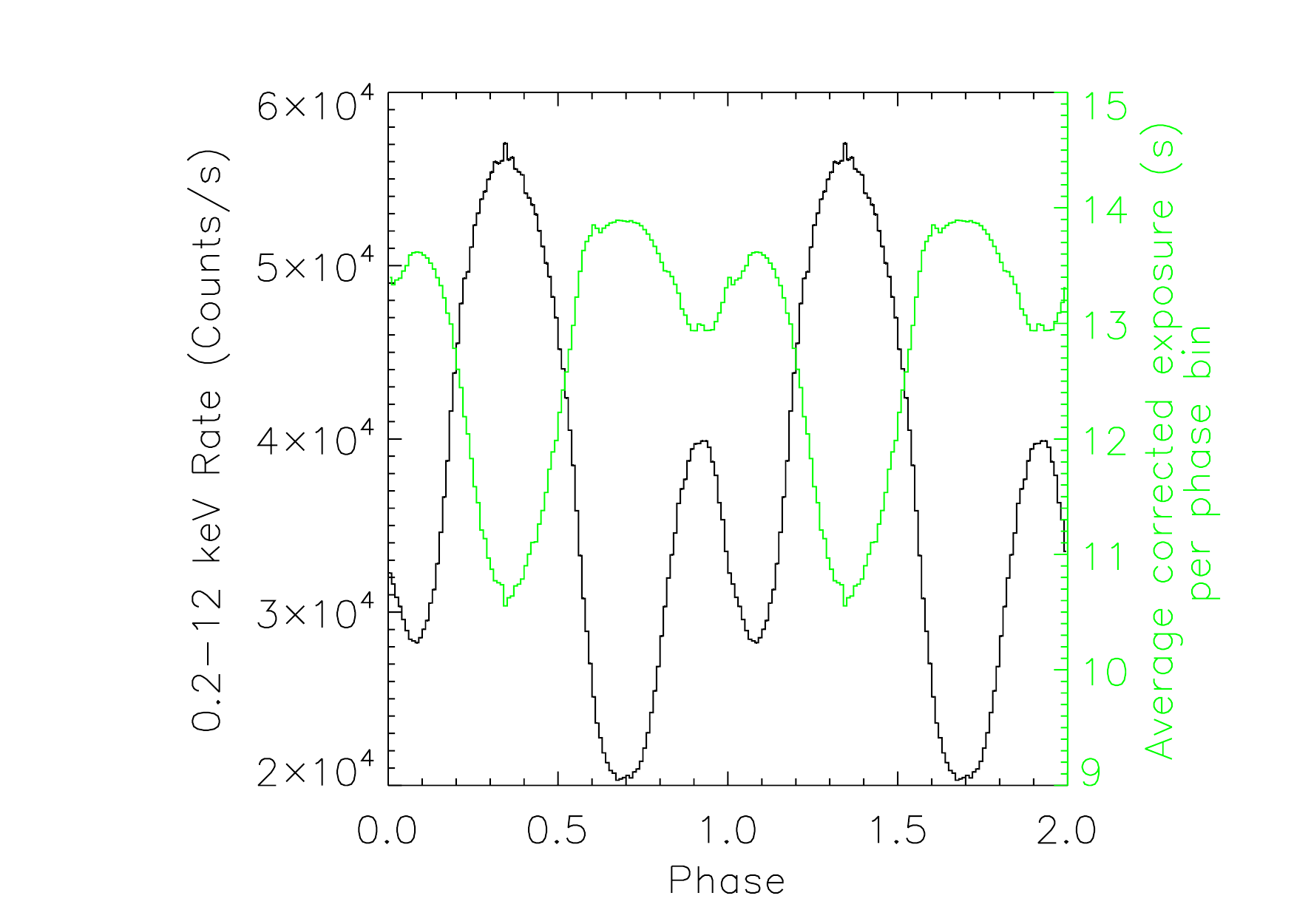}	
	\end{center}
\caption{Swift~J0243.6+6124 folded pulse profile (two cycles plotted) of one of the brightest \textit{NICER} observations, ObsID \# 1050390115, on 2017 November 7. The black histogram and left-hand y-axis denote the folded pulse profile in the 0.2--12 keV band, averaged over the 7 MPU slices, after corrections for GTI exposure and deadtime have been applied. The green curve and right-hand y-axis denote the deadtime corrected exposure per phase bin, averaged over the 7 MPU slices for this observation. The total deadtime corrected exposure, averaged over MPU slices, for this observation is 1276.5 s. If the bin-to-bin variations were not present, each bin would have an exposure of 12.765 s. The exposure per bin differs for each observation and shows bin-to-bin variations $>$1\% compared to uniform exposure for total count rates $> 2000$ counts s$^{-1}$.  \label{fig:gtiexposure}} 
\end{figure}

Second, the GTI exposure for each individual phase bin was then computed, again separately for each MPU slice. These good time intervals were taken from the GTI extension of the calibrated event files. In the brightest observations, shown in Figure~\ref{fig:gtiexposure}, the exposure time per phase bin varies dramatically from bin-to-bin. This is caused primarily by buffers filling up at each MPU slice, which results in very short GTIs. To account for these bin-to-bin exposure time variations, the GTIs longer than a phase bin were divided into pieces 1/10 of a phase bin long. Their start times were corrected for the pulsar's orbit, a pulse phase was computed for each time, and the total exposure was accumulated for each phase bin. GTIs shorter than a phase bin were corrected and accumulated without slicing.  

Third, deadtimes were obtained from the unfiltered event files in order to include all events that could contribute to deadtime, both the good events included in the pulse profile, and the non-source events filtered out of the profile. Deadtime for each event is stored in the \textit{NICER} fits files as a value from 0--127. Each of these values $V_{\rm{dead}}$ corresponds to a range of 16 clock ticks, i.e., 0 is 0--15 ticks, 1 is 16--31, and so on, with a maximum recorded dead time of 127 or 2032 ticks. The dead time is computed in seconds as $t_{\rm{dead}} = (V_{\rm{dead}}+0.5) \times 16/(25.8 \times 10^6$ Hz), where the factor of 0.5 shifts to the center of the recorded interval, 16 is the number of ticks per interval, and 25.8 MHz is the clock frequency. For each event, the event times were corrected for the pulsar's orbit, phases were calculated, and the dead time was accumulated in each phase bin. Like the folded events and folded GTI intervals, the deadtimes were accumulated for each MPU slice separately. 

Fourth, to obtain the deadtime corrected exposure time for each phase bin, $\epsilon$, the total deadtime for each phase bin was subtracted from the GTI exposure for each bin. This resulted in an exposure time for each of 100 phase bins for each MPU slice. Table~\ref{table:nicerobs} lists the total deadtime corrected exposure averaged over the seven MPU slices for each observation. The exposure is typically the same for all MPU slices to within 1\% for average count rates $< 1000$ counts/s. For the highest count rates, the exposure can vary by as much as 10\% from MPU slice to MPU slice. 

Finally, the count rate in each phase bin was obtained by simply dividing the total number of filtered events in each phase bin by the deadtime corrected exposure in each phase bin. Poisson-distributed uncertainties were assumed, and were calculated as the square root of the counts divided by the deadtime corrected exposure in each bin. The last step was to combine the count rate pulse profiles for each MPU slice into a single \textit{NICER} pulse profile for each observation. Because 52 of 56 \textit{NICER} detectors are operating on orbit, the MPU slices have differing numbers of inputs. The count rate profiles were corrected by a simple factor so that all would have the expected rate for eight detectors. The profiles from MPU slices 1 and 6 were multiplied by $8/7$, and the profile from MPU slice 2 was multiplied by $8/6$, then the profiles from all seven MPU slices were added together and their errors were combined in quadrature. 

\subsection{\textit{NICER} Pulse Profile Analysis}\label{subsec:nicerpulse}

\textit{NICER} observations of Swift J0243.6+6124 began early in the outburst on 2017 Oct 2, triggered by an alert from \textit{MAXI} \citep{Sugita_2017} on 2017 September 29. 
%The outburst profile is asymmetric, with a faster rise than fall as shown in Figure~\ref{fig:lcandprofiles}, as is common for giant outbursts in Be/X-ray binaries \citep{Reig+13}. Inset plots in Figure~\ref{fig:lcandprofiles} show pulse profile evolution throughout the outburst. 
%% The pulse shapes are similar at similar count rates both in the rise and fall of the outburst, evolving from highly 
%% complex at lower count rates, to single peaked, to double peaked and nearly symmetric, and to double peaked. 
The outburst profile is asymmetric, with a faster rise than fall as shown in Figure~\ref{fig:lcandprofiles}, as is common for giant outbursts in Be/X-ray binaries \citep{Reig+13}. Inset plots in Figure~\ref{fig:lcandprofiles} show pulse profile evolution  in the full 0.2--12 keV band throughout the outburst. The pulse shapes are similar at similar count rates both in the rise and fall of the outburst, evolving from highly complex  (Fig.~\ref{fig:lcandprofiles} insets a \& h) at lower count rates, to single peaked  (inset b \& g), to double peaked and nearly symmetric  (inset c \& f), and to double peaked but asymmetric at the outburst peak  (inset d \& e). Detailed energy dependent pulse profiles for each \textit{NICER} observation are shown in the Appendix.

 \textit{NICER} observations continued as the outburst faded. The flux reached a minimum in the \textit{NICER} band on MJD 58204 (See Table 1). Pulsations were still detected and the 0.2--12 keV profile was a simple symmetric single peak, similar to MJD 58176, the last profile shown in Figure~\ref{fig:A1} in the Appendix. Then the flux again began to increase and a normal (type I) outburst was detected with \textit{NICER}, GBM, and \textit{Swift} BAT peaking on MJD 58220. A second normal outburst peaking near MJD 58247 was detected with GBM and \textit{Swift} BAT, but was not observed with \textit{NICER}. 

%Energy dependent pulse profiles were generated by selecting events in the 0.2--1, 1--2, 2--3, 3--5, 5--8, and 8--12 keV bands. Figures~\ref{fig:energy_profiles_start}-e show the evolution of the pulse profiles with time and with energy. Error bars are standard 1-$\sigma$ errors assuming Poisson statistics and are smaller than the line thickness in the brighter observations. Profiles are not phase connected. At lower average count rates ($<1000$ cps), early and late in the outburst, considerable energy dependent pulse shape variations are observed and the pulse profile is very complex, with the largest peak closest to the deepest minimum.
%Then a transition occurs at 500--1000 cps where the pulse profile becomes primarily a single asymmetric peak and seems to reverse in phase as to which peak is dominant. Above about 3000 cps, the profile becomes more symmetric and gradually splits into two equal peaks and the energy dependence becomes less dramatic. The minimum deepens between the two peaks as the intensity increases. As the outburst approaches its peak $>16000$ cps, the profile again becomes asymmetric with the peak closer to the deeper minimum again dominating. As the outburst decays, the profiles evolve very similarly through the shapes and complexities observed during the outburst rise.   This analysis demonstrates the wealth of observations made possible by \textit{NICER} for an extremely bright source.

\begin{figure}\label{fig:pulse_fraction}
\begin{center}
\includegraphics[width=0.475\textwidth]{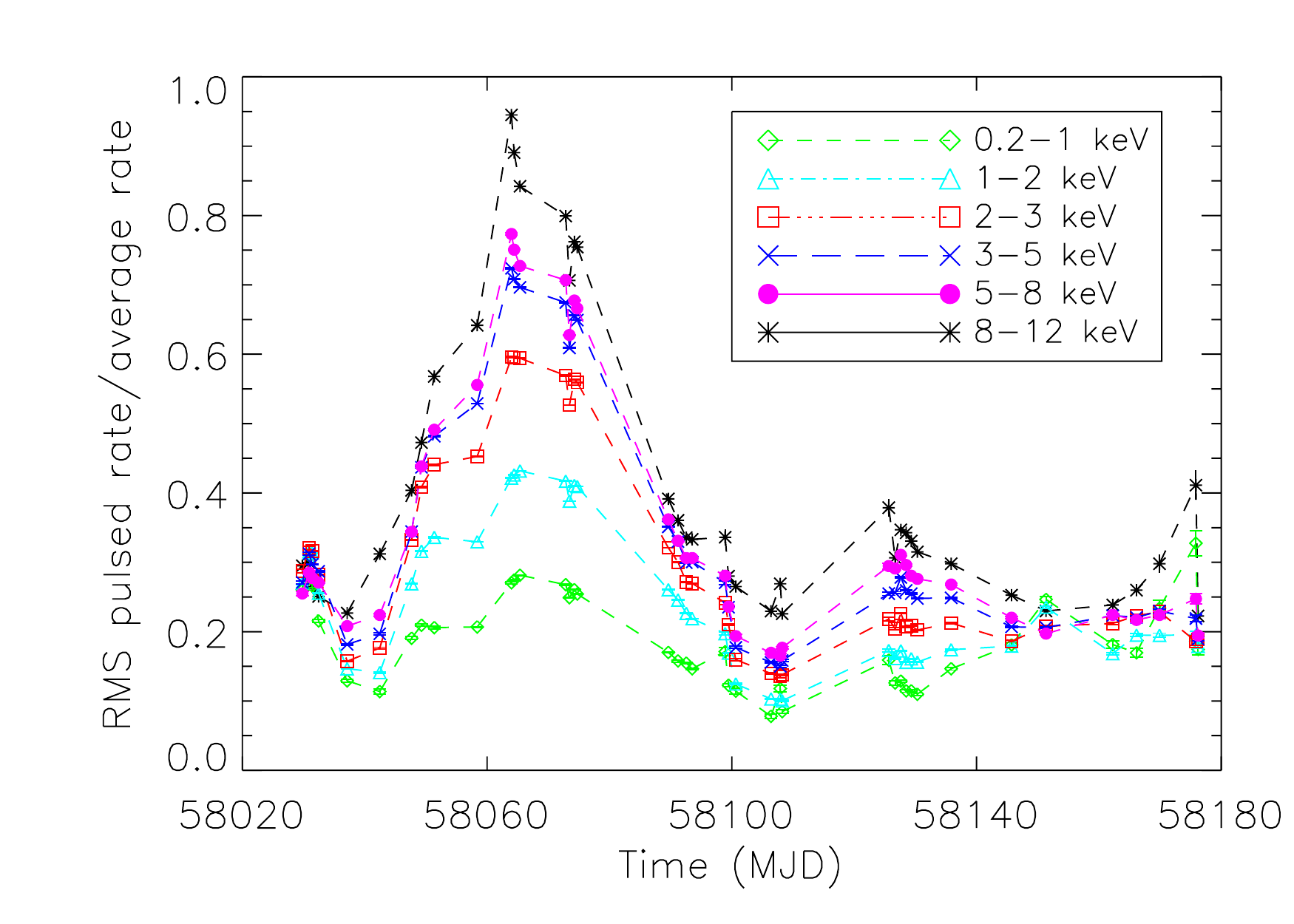}
\includegraphics[width=0.475\textwidth]
{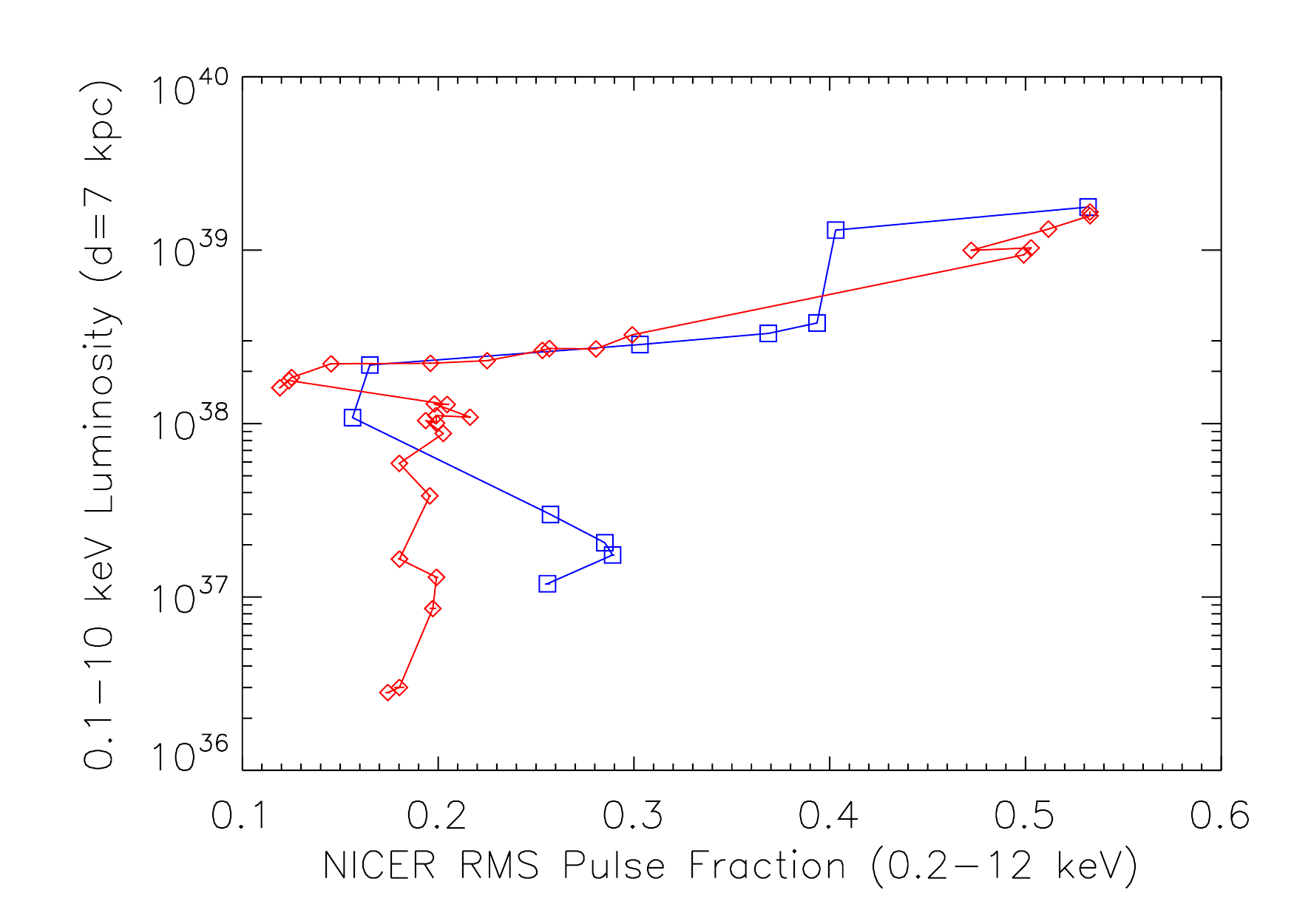}
\end{center}
\caption{(Left):Swift J0243.6+6124 root-mean-squared pulsed fraction in several energy bands versus time from \textit{NICER} observations.  (Right): Swift J0243.6+6124 0.1--10 keV luminosity (d=7 kpc) vs. root-mean-squared pulsed fraction (0.2--12 keV). The rising portion of the outburst and the fading portion are denoted by blue squares and red diamonds, respectively.  During the main peak of the outburst, the pulsed fraction generally increases with increasing count rate and increasing energy, especially after a striking turn-around near $L > 10^{38} (d/7\,{\rm kpc})^2$. Error bars are smaller than the plotted symbols in both panels.}
\label{fig:nicer_fraction}
\end{figure}

% \setcounter{figure}{4}
The root-mean-squared pulsed fraction $f_{\rm rms}$ was computed for the full 0.2--12 keV band and for each of six energy bands: 0.2--1, 1--2, 2--3, 3--5, 5--8, and 8--12 keV. The pulsed fraction is given by 
\begin{equation}
\label{eq:rms}
f_{\rm rms} = \frac{(\Sigma^{N}_{j=1} (r_j - \bar{r})^2/N)^{1/2}}{\bar{r}}
\end{equation}
where $r_j$ is the count rate in phase bin $j$ and $\bar{r}$ is the phase-averaged count rate. The time history of the pulsed fraction for the six energy bands is shown in Figure~\ref{fig:nicer_fraction} (left panel). 
%Early in the outburst, when the pulse profiles are the most complex, the 0.2-12 keV pulse fraction is higher, at about 20\%. Then, when the pulse profile simplifies to a single peak, the 0.2-12 keV pulse fraction also is lower at about 10\%. As the overall outburst intensity increases, the 0.2-12 keV pulse fraction increases as well, peaking at about 38\% at the outburst peak. As the outburst declines, the 0.2-12 keV pulse fraction decreases, again dropping to about 10\% when the profiles transition to a single peaked shape. The pulse fraction increases as the count rate declines further and the profile shape becomes more complex, but not quite to the level of the earliest observations. 
The time evolution of $f_{\rm rms}$ in the individual energy bands is similar to the full-band $f_{\rm rms}$. However, the peak pulsed fraction increases substantially with energy.  The relationship between the full 0.2--12 keV band $f_{\rm rms}$ and the 0.1--10 keV luminosity (see Table 1) is shown in Figure~\ref{fig:nicer_fraction} (right panel). A clear turn-around is present in both the rising and fading portions of the outburst.

\subsection{\textit{NICER} Power Spectra}\label{subsec:powspec}
We analyzed variability on time-scales other than the neutron star
spin period by estimating the power spectral density (PSD; $P(f)$) for
each epoch.  First, we binned the data in 100\,ms bins using photon
times that had been corrected to the solar system barycenter.  We computed deadtime and exposure, $\epsilon_i$, as above, allowing an estimate
of the true count rate, $r_i=c_i/\epsilon_i$.

Because \textit{NICER} only views the source for a fraction of each
ISS orbit, the window function for a given epoch is typically complicated, leading to a convolution of the true power spectrum with a series
of $\mathrm{sinc}$ functions.  In the PSD, this is squared, so a pure tone
will be spread into side lobes whose magnitude decreases as $1/f^2$, a phenomenon known as spectral leakage.  It is
therefore critical both to eliminate as much pulsed power from the
signal as possible and to pay careful attention to the resulting
spectra for evidence of such leakage.

To reduce the pulsed signal, we first estimate the spin period and
period derivative for each epoch, taking into account binary motion,
and use this to assign a pulse phase to each time bin, $\phi(t_i)$.
We then approximate the power at the spin frequency with a truncated
Fourier series whose coefficients we estimate as, e.g., $\sum_i r_i
\cos(2\pi k\phi_i) / \sum_i r_i$, with $k$ between 1 and 10.  We
evaluate the Fourier series at each time bin using $\phi(t_i)$ and $\epsilon_i$ and
subtract this from the observed rate.

To compute a PSD for visualization for each epoch, we selected contiguous segments of data with at least 2048 samples, then zero-padded the (mean subtracted) data to $2^{14}$ samples.  Most segments were shorter than this.  We computed the discrete Fourier transform using the FFT algorithm and co-added the resulting spectra using the interval lengths as weights. We normalized the spectra according to the convention common
in the literature \citep[e.g.,][]{Nowak99}, in which the PSD integrated
over frequency yields the rate-normalized variance, $f_{\rm rms}^2$ (see
Equation \ref{eq:rms}).  Using shorter, contiguous segments degrades spectral resolution and precludes examination of frequencies $<$1\,mHz, but it provides superior window functions and is more robust against any uncorrected deadtime effects that evolve over an epoch.

\begin{figure}
\begin{center}
\includegraphics[width=\textwidth]{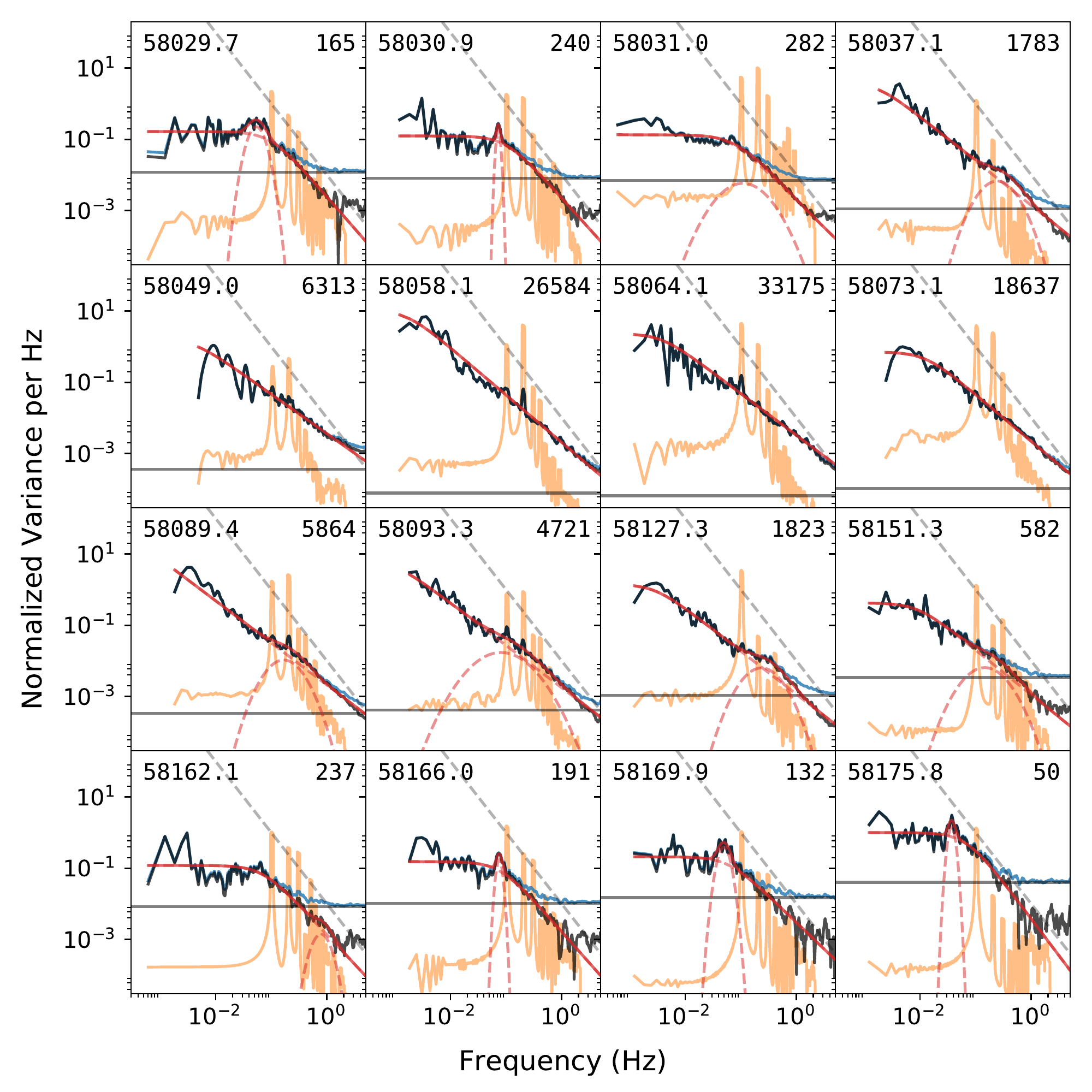}
\end{center}
\caption{The power spectral densities for 16 epochs with \textit{NICER}.  The MJD and count rate (from Table \ref{table:nicerobs}) are given for each observation.  The main features are described in the text, and are briefly: light tan, estimated pulsed PSD; blue, smoothed PSD; dark horizontal line, white noise level; black, smoothed and noise-subtracted PSD; red, total (solid) and component (dashed) fits; dashed gray line, a $1/f^2$ line for visual comparison. }
\label{fig:powspec}
\end{figure}

The resulting PSDs for 16 epochs appear in Figure \ref{fig:powspec}, which we
explain in detail here.  For display purposes, all spectra have been logarithmically smoothed with a resolution of about 0.05 dB.  The tan trace in each panel of Figure \ref{fig:powspec}, with an obvious harmonic
series, is the PSD of the mean pulsed signal, computed as described above,
multiplied by the exposure for the particular observation.  The fundamental appears at the spin frequency 0.1014\,Hz.  It is
typically quite strong compared to the broadband variations.  Because
the exposure (window function) is the same as for the real data, this
PSD is useful for assessing the effect of the window function on pure
tones.   We also include a dashed gray line whose slope is fixed to $1/f^2$ to aid in assessing spectral indices and leakage.  The blue trace shows the PSD of the observed count rate after
subtracting the mean pulsed component.  Some remaining pulsed power is
evident, e.g., in the observation on MJD 58031.0, but such power is small compared to the broadband
features.  The dark horizontal line shows the white noise level, which for Poisson statistics is given by $<$$r/\epsilon$$>$.  We find this to be in good agreement with the observed spectra.   Finally, the black trace shows the power spectrum with white noise subtracted.  This is the best empirical estimate of the aperiodic PSD and can be used for visual characterization.  We note here that we also computed PSDs by omitting each MPU slice in turn, and found no substantial difference.  Finally, we note that the relative power levels are high, such that any contributions from, e.g., pointing jitter, are negligible by comparison.

We have also fit functional forms to the power spectra using a maximum
likelihood approach.  For this purpose, we did not co-add the segments within an epoch, but rather computed FFTs with lengths exactly equal to the segments, such that the resulting FFT bins are statistically independent.  The elements of the PSD then follow a $\chi^2_2$ distribution when scaled
appropriately, so the log likelihood becomes $\log\mathcal{L}=-\sum_i
\log P_m(f_i,\lambda) + P(f_i)/P_m(f_i,\lambda)$, where $P_m(f)$ is
the model PSD for parameters $\lambda$, and we maximize this
likelihood evaluated over the unsmoothed power
spectra.  We exclude spectral bins with $f>2$\,Hz to minimize the effects of aliased power from $f>5$\,Hz (the Nyquist frequency).  In the literature, lorentzians of various widths are widely
used to model PSDs \citep[e.g.,][]{Reig+13}, but they asymptotically follow
$1/f^2$.  The observed PSDs are typically shallower than this, so we
have instead fit the broadband power with a power law with a low
frequency cutoff, $P(f)\propto[1+(f/f_c)^2]^{\alpha/2}$.  We model
narrower features with a lognormal function, and we include the white noise contribution given above.  The resulting power
law fits are shown with red traces, individual components appearing as dashed lines and the total model solid.

There is clear evolution in the power spectra.  The earliest observation is essentially white at low frequencies with a low quality QPO-like feature at 50\,mHz. 
QPOs have been observed already in a number of accreting pulsars ranging from $\sim1\,$mHz to $\sim40\,$Hz (see, e.g., \citealt{Paul2011,Reig+13}), and are generally associated with inhomogeneities in the inner accretion disk.
%\citep{Paul+Rao98,James+10, Dugair+13}
The spectral break is almost exactly at the neutron star spin frequency ($0.1014\,$Hz), and the high-frequency behavior is slightly shallower than $1/f^2$.  Inspection of the window function shows this steeper tail cannot simply be spectral leakage from the QPO, so it reflects a distinct physical process.  On MJD 58030.9, the QPO is stronger, narrower, and at higher frequency, about 70\,mHz.  However, because of the low spectral resolution at these frequencies, we are unable to quantitatively characterize any evolution of the QPO frequency and shape with luminosity.

As the luminosity increases, the power law cutoff disappears and the power law becomes harder with typical indices of $-1.1$ to $-1.3$.  At higher frequencies, a broad peak of power, about a decade wide and largely overlapping the band including the spin frequency and its harmonics, appears (MJD 58031.0--58049.0).  At these epochs, this broad peak can be identified with the lognormal component (dashed red lines) of the model fit.  Interestingly, this component is not necessarily new, as the normalized power at $\sim$0.5\,Hz is almost constant at all epochs; thus, this component may instead model a deficit in power around 0.1\,Hz.  In the brightest part of the outburst (MJD 58049.0--58073.1, the spectra are consistent with a single power law.  As the luminosity decays, the broad high frequency peak re-appears (MJD 58089.4--58151.3).  Finally, the spectrum returns to a flat low frequency pedestal terminated by a cutoff/QPO near the spin frequency and a steep ($\sim$$-1.7$ to $-2.0$) high-frequency tail.

\subsection{\textit{NICER} hardness ratios}\label{subsec:hardness}

In many accreting X-ray pulsars the source spectrum is found to be flux-dependent.
When the observed spectrum is fit using a power-law component, the spectral photon index $\Gamma$ proportionally or inversely correlates with the source luminosity, according to the accretion regime.
A critical luminosity $L_c$ distinguishes between two accretion regimes \citep{Basko+Sunyaev76}: at relatively low luminosities $L<L_c$, the height of the accretion structure is governed by the ram pressure of the infalling material, called ``sub-critical'', while at higher luminosities $L>L_c$ a radiation-dominated shock rises in the magnetic column, governed by radiation pressure in the accreting material, called ``super-critical''.
For sub-critical sources the spectrum of many accreting X-ray pulsars has been found to harden as the observed X-ray flux increases 
\citep{Klochkov+11, Fuerst+14, Malacaria+15, Postnov+15, Epili+17}.
On the other hand, super-critical sources show the opposite behavior, that is a spectral softening as the flux increases (\citealt{Reig+13,Postnov+15,Epili+17}).
Therefore, thanks to the opposite observed behaviors, the flux dependence of the spectral photon index can be used to identify the accretion regime.

A model-independent way to probe whether a source is accreting at sub- or super-critical regime is through the analysis of the hardness ratio.
\citet{Reig+13} studied the patterns of variability of the hardness ratio during outbursts in a number of Be/X-ray pulsars.
Those authors identified two branches, the horizontal branch (HB) and the diagonal branch (DB), in the hardness-intensity diagrams for these sources. The spectral hardness follows a certain path in the hardness-intensity diagram and undergoes state transitions from one branch to another, that is, it ``turns over" if the source luminosity passes through a certain critical value.
%which, in all cases, was close to the source critical luminosity $L_c$.Thus, turn-overs in the hardness/flux dependences most likely reflect a transition between the sub- and super-critical regimes.% (see Fig.~\ref{fig:soft_color}).
These diagrams used a intensity in the $4-30\,$keV band, using data from the \textit{RXTE} Proportional Counter Array (PCA, \citealt{Jahoda_1996}) and a soft color defined as the ratio between the count rates in two energy bands, namely $7-10\,$keV and $4-7\,$keV. 

\begin{figure}[ht!]
\plotone{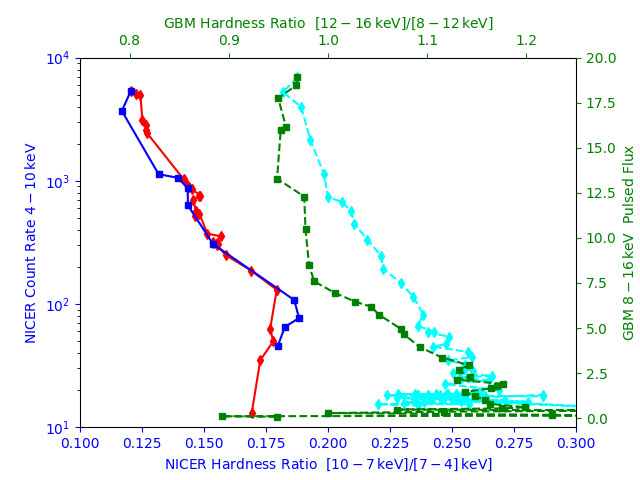}
\caption{Hardness ratio evolution with flux for Swift~J0243.6+6124. Squares show the rise stage of the outburst, and diamonds the decay. Both \textit{NICER} (blue squares and red diamonds connected by a solid line) and \textit{Fermi} GBM (green squares and cyan diamonds connected by dashed line segments) hardness ratios are shown.
Error bars are smaller than the plotted symbols.
Note the different scales on the y-axis: logarithmic for \textit{NICER} counts, linear for GBM pulsed fluxes.
Spectral evolution and hysteresis at high luminosity are evident.}\label{fig:soft_color}
\end{figure}

We investigated the spectral behavior of Swift~J0243.6+6124 by examining the hardness ratio evolution throughout the entire outburst, with both \textit{NICER} and \textit{Fermi} GBM data. 
\textit{NICER} data are useful for reproducing the original \citet{Reig+13}'s soft color (SC, $7-10\,$keV/$4-7\,$keV), while GBM enables one to explore the hardness ratio behavior at higher energy and offers uniformly sampled coverage of the entire outburst.
The $4-10\,$keV \textit{NICER} count rate has been used as a proxy for the $4-30\,$keV total intensity originally employed by \citet{Reig+13}.
For GBM, the hardness ratio has been chosen as the ratio between the pulsed flux levels in the $12-16$ and $8-12\,$keV energy ranges, whose edges are defined by the channel edges of CSPEC data (See Section~\ref{subsec:GBM_instrument}), while the chosen bands are such that the intervals are insensitive to interstellar absorption while guaranteeing sufficient statistics.
Likewise, the $8-16\,$keV pulsed flux has been used to represent the GBM outburst intensity.
Fig.~\ref{fig:soft_color} shows the evolution of the hardness ratios with intensity for both the \textit{NICER} and the GBM energy bands.
GBM data clearly show both the HB at lower intensity and the DB at higher intensity, with a turn-over taking place in between.
However, \textit{NICER} seems to trace the DB more clearly, while only a hint of the HB appears as the luminosity decreases along the decay stage of the outburst. 
Moreover, the turn-over traced in the hardness-intensity diagram by \textit{NICER} data seems to happen at a lower luminosity than GBM's.
We will discuss possible implications of these results in Sect.~\ref{subsec:hardness_discussion}.

%\begin{figure}
%\includegraphics[width=\textwidth]{ratio_vs_flux_reig_bands.pdf}
%\caption{Soft color (SC, 7-10 keV/4-7 keV) versus intensity in the 4-10 keV band from NICER data. Red squares show the outburst rise while blue diamonds show the outburst decay. The first NICER observations are on the bottom right, forming a possible small horizontal branch. Time evolves from the bottom right to the top left and then back again. \label{nicer_hardness}}
%\end{figure}

\subsection{\textit{NICER} spectral analysis}\label{subsec:spectral}

%{\it
The energy spectra of accretion powered X-ray pulsars can be described by simple phenomenological models despite the complex emission processes acting close to the neutron star. In support of our timing studies and to explore the nature of X-ray emission, we attempted to understand the spectral behavior of Swift~J0243.6+6124 during the outburst. Photon energy spectra were extracted from all available  data (see Table~\ref{table:nicerobs}). We followed the standard analysis method as described in Sections~\ref{subsec:nicerxti} and \ref{subsec:nicerdeadtime}. In addition to conventional filtering criteria, a {\tt SUNSHINE} flag was also added to GTI that can separate the data 
at times when \textit{NICER} was exposed to direct sunlight (``day'' orbits), and times when \textit{NICER} was shadowed by the Earth (``night'' orbits). This was done as a first step to address the known instrument artifacts from O, Si and Au in the \textit{NICER} spectra at energies close to 0.5, 1.8 and 2.2~keV, respectively (see also \citet{Ludlam2018}). These residuals were  more prominent in high count rate observations, complicating spectral analysis at present for brighter sources. 

%\textbf{CM: It might be useful to address the reader about the Crab normalization technique a bit more in detail. If Renee's paper has been submitted by the time we submit, we could refer to that paper in this paragraph too (same as the paragraph above). ... 
%GKJ: Hey... I am in process of updating texts. I do have plan to include all these information.}

We adopted a normalization technique that uses the Crab Nebula's observed spectrum to mitigate these instrument features. For this, we extracted Crab spectra from (ObsIDs 1011010101, 1011010201, and  1013010101-1013010122), in a manner similar to Swift J0243.6+6124. The products from day and night orbits were combined into respective groups using the FTOOL {\tt addspec}. We accumulated time-averaged spectra of the Crab with an effective exposure of $\sim$26 ks for day and $\sim$14~ks for night orbits. Together with a background dataset (\textit{RXTE} blank sky field 6; \citealt{Jahoda2006}), and the instrument response (energy redistribution matrix and effective area files, version 0.06), the Crab spectrum was described in the 0.25--10~keV range by an absorbed power-law. A poor fit was obtained at this point due to residuals in the soft X-rays arising from the preliminary instrument calibration at this stage. We froze the absorption column density from the above fit and refitted the continuum in a flat, feature-less (4--10 keV) range. At these energies the model yielded a better fit with a photon index close to 2. We next extrapolated the energy spectrum back down to 0.25~keV and created a fake spectrum using the {\tt fakeit} command in {\tt XSPEC} for  the Crab exposure. By construction, simulated data was free of significant instrumental bias below 4~keV, and taking its ratio with the Crab spectrum, therefore, generated a  ``template'' that carried the artifacts. Finally we divided the Swift~J0243.6+6124 spectrum by the template, using the FTOOL {\tt MATHPHA}. 

The 3--79 keV energy spectrum of the pulsar with {\it NuSTAR} can be expressed by a cutoff power-law, high energy cutoff power-law, and/or negative and positive exponential cutoff power-law models modified with a blackbody component \citep{Jaisawal_2018}. For the purpose of estimating the pulsar's unabsorbed flux from \textit{NICER} spectral analysis, we instead considered a simple cutoff power-law  to fit the  0.3--10 keV continuum. We noted that spectra at brighter phases of the outburst ($>$3 Crab intensity) suffered from instrument features, despite the Crab-renormalization technique, limiting the present study. Nevertheless, our results show a remarkable intensity variation and gradual spectral evolution across the outburst consistent with the 
accretion-regime transition scenario outlined in Sect.~\ref{subsec:hardness} and discussed in greater detail in Sect.~\ref{sec:discuss}. The unabsorbed flux has been approximately corrected for deadtime as derived in Section~\ref{subsec:nicerdeadtime} for each observation.  The 0.1--10 keV luminosity was calculated by assuming isotropic emission and  a source distance of 7~kpc \citep{Bailer-Jones2018} and is presented in Table~\ref{table:nicerobs}. A detailed spectral analysis of the pulsar will be presented in a future paper using refined calibration, response, and effective area files. 
%}

\section{GBM Observations of Swift J0243.6+6124}\label{sec:gbm_analysis}

\subsection{GBM Orbital analysis}\label{subsec:gbmorbit}

We have determined an orbital ephemeris for Swift J0243.6+6124 using pulse frequencies from GBM and the 15--50 keV count rates from \textit{Swift} BAT. We binned the GBM Channel 1 (12--25 keV) CTIME data from MJD 58027--58150 to 250 ms and then fitted it to a semi-empirical background model.  Next we subtracted the background model and extracted pulsed flux and frequencies for Swift J0243.6+6124 in the manner described in \citet{Finger_1999, Jenke_2012}.  We performed a search for frequency and frequency derivative using pulse profiles folded over one day intervals.
The intrinsic spin--up of an accreting HMXB pulsar, $\dot{\nu}$, is proportional to $L^{6/7}$ when accretion is mediated through a disk \citep{Lamb1973}.  We applied the techniques described in \citep{Tsygankov2016} to model this intrinsic spin--up using the \textit{Swift} BAT rates as a proxy for the X-ray luminosity, because simple polynomial models failed to fully model the observed spin-up.

The BAT had considerable issues with measuring the rates for this source between MJD 58060 and 58075. This was caused by a known data overflow problem at very high count rates. When the counts exceeded the maximum value that could be stored, they wrapped and were instead stored as extremely low values\footnote{\url{https://swift.gsfc.nasa.gov/results/transients/weak/SwiftJ0243.6p6124/}}.  In order to correct the rates, only the peak rates during this period were used and the rates and errors corresponding to the times of measured frequencies were interpolated.  The model emission frequencies were determined by integrating the modeled spin-up determened from the \textit{Swift} BAT rates.  The model frequencies were corrected for the current orbital parameters as described in \citep{Deeter1981}.  The model frequencies were then compared to the measured frequencies.  Using the Metropolis-Hastings Algorithm \citep{Hastings70}, a search was performed to find the best-fit orbital parameters.  The torque model based on the \textit{Swift} BAT rates failed to produce a reasonable fit to the data due to the high spin-up during the peak of the outburst.  Parameters that increased the index on the BAT rates at peak flux were added to the torque model: one parameter that increased the power index on the BAT rates above a threshold and, $\beta$, the threshold on the BAT rates in which the addition to the index would be applied.  The new model for the spin-up rate was:  
\begin{eqnarray}
\dot{\nu} \propto R^{i_1 + i_2}
\end{eqnarray}
where $R$ are the interpolated BAT rates, $i_{1}$ is the index when the BAT rates are below $\beta$ and $i_{1} + i_{2}$ is the index when the BAT rates are above $\beta$.

The new model produced a better fit to the data but there were still significant residuals remaining.  Using the best-fit parameters to the orbit, the measured frequencies where corrected for the best-fit orbit and a detailed phase model was determined.  Using the new phase model, a new search for pulsations was performed.  A phase offset $\delta \phi$ from the phase model was estimated for each 1 day mean pulse profile by fitting the Fourier amplitudes of the profile to a scaled and shifted template profile.  The new pulsar orbital phase offsets were fit to the orbital model given by \cite{Deeter_1981} using circular orbital elements an a polynomial background model to remove the remaining residuals.  Only the tail of the outburst (between MJD 58098 and 58154) was used to avoid the large torque noise during the peak of the outburst.   The refined orbital elements were used to further refine the phase model and again a new search for pulsations was performed.  This was repeated until there was no significant change in the best fit orbital elements.  To quote orbital element uncertainties representative of systematic effects, statistical errors in the frequency measurements were inflated to obtain a reduced chi-squared near 1.0 (46.6/43 dof). Sources of systematic error may include changes in the emission beam within the integration interval or throughout the outburst.

\begin{table*}[h!]
\begin{center}
\caption{Orbital Ephemeris for Swift J0243.6+6124.}
\begin{tabular}{lcc}
\tableline\tableline

$P_{\rm orb}$        &   27.587 $\pm$ 0.016  & days   \\  
$T_{\pi/2}$                     &   58115.6640  $\pm$  0.0049 & MJD   \\
$a_{\rm x}\sin i$                   &   115.84  $\pm$ 0.32 & lt-sec     \\
$e$             &    0.09848  $\pm$  0.00042 &     \\
$\omega$  &   286.44  $\pm$  0.16 & degrees      \\
$f(M)$             &   2.138 $\pm$  0.015 & M$_\odot$    \\
$\chi^{2}$                      &    46.6 / 43 dof &              \\

\tableline
\end{tabular}
\end{center}
\end{table*}

The frequency history, pulsed flux and ephemeris for this source and all other sources monitored by the GBM Pulsar Monitoring team may be found at the GBM pulsar website\footnote{\url{http://gammaray.msfc.nasa.gov/gbm/science/pulsars.html}}.  See Appendix A for a detailed analysis of GBM energy dependent pulse profiles and their comparison with \textit{NICER}. 

%\subsection{GBM Pulse Profile analysis}\label{subsec:gbmpulseanalysis}

%Pulse profiles were generated in nine energy bands from 5--100 keV using GBM CTTE data. For each day of data, harmonic expansions including 24 sine and cosine terms in a pulse phase model were fitted to the GBM count rates. Times were barycentered and corrected for the orbital ephemeris described in the previous section. These harmonic expansions were converted to a more typical pulse profile and are plotted in Figure~\ref{fig:gbm_energy_profiles_end}a-e. Error bars are plotted for every $\sim 4$ bins, corresponding to the independent harmonics. Each profile has been shifted so that the minimum in the 8--12 keV band is at phase zero. Comparing these profiles to those in Figure~\ref{fig:energy_profiles_start}-e, the higher energy profiles also evolve from complex at first to asymmetric and single peaked, but narrower, to double peaked, but more asymmetric than at lower energies, back to single  peaked, and complex. When the profiles are the most complex, the phasing of the GBM and \textit{NICER} profiles may not be consistent. 

%\setcounter{figure}{8}

\section{Discussion}\label{sec:discuss}

\subsection{Pulse profile evolution at different accretion regimes}\label{subsec:regimes}

The amount of data that \textit{NICER}, \textit{Fermi} GBM and \textit{Swift} have accumulated on the outburst of Swift J0243.6+6124 is very large and it is not easy to understand every detail in light of our current understanding of the accretion flows onto magnetic neutron stars. That the neutron star is magnetic is clear from the fact that we see distinct pulses at a discrete frequency with clear indications of spin period changes over the course of the outburst. And further, those spin period changes appear to be correlated with the luminosity (thus the accretion rate onto the neutron star) lending confirmation to the idea that we are seeing rapid spin-up of a neutron star in a close orbit around a Be early-type star.
The observation of a spin-up torque that is correlated with the luminosity is a clear indication that the accretion is mediated through an accretion disk. More quantitatively, an accretion disk will form around the pulsar if the matter in the Be-star's circumstellar disk has a net angular momentum greater than the angular momentum of matter being spun around by the rotating magnetosphere. We can calculate the radius of the Be star's circumstellar disk  $R_{\rm disk}$ using the relation from \citet{Hanuschik1989} that relates it to the measured  H$\alpha$ equivalent width, $EW=-10.3\mbox{\AA}$ from \citet{Kouroubatzakis_2017}, to obtain $R_{\rm disk}\approx 11 R_{\rm c}$.  Then using this value in the criterion from \citet{Klus2014} (see also \citealt{Shapiro1976,Wang1981}) along with parameters of the system (i.e., companion mass and radius of $\sim 16M_\odot$ and $\sim 7R_\odot$, respectively, from \citet{Bikmaev+17}), a range of mass accretion rates $\dot{M}\sim 10^{-11}-10^{-8}M_\odot$~yr$^{-1}$, and an assumed magnetic field for the neutron star in the range $B\sim 10^{12}-10^{14}$~G, we find that matter accreted onto the pulsar from the circumstellar disk of the Be star first forms an accretion disk around the pulsar.

Early in the outburst, before approximately MJD~58033, the pulse profile is complex, but dominated by a single asymmetric peak (see Fig.~\ref{fig:lcandprofiles} inset a). 
A remarkable change occurs near MJD~58037, in that the pulse shape transitions to a single simpler peak, both in the \textit{NICER}  (Fig.~\ref{fig:lcandprofiles} inset b) and GBM energy bands (See Appendix, Fig.~\ref{fig:A1}). 
Then, as the accretion luminosity takes off, a clear two-peak structure emerges (Fig.~\ref{fig:lcandprofiles} inset c) and dominates the emission pattern at all energy ranges (See also Fig.~\ref{fig:A3}). 
This two-peak flow arrangement lasts through the peak luminosity of $ \sim1.8\times 10^{39}$ erg s$^{-1}$ at MJD~58064  (Fig.~\ref{fig:lcandprofiles} insets d, e, \& f) and as the luminosity goes down below $ \sim 2 \times 10^{38}$ erg s$^{-1}$. Once MJD~58100 is past, both the \textit{NICER}  (Fig.~\ref{fig:lcandprofiles} inset g) and GBM (Fig.~\ref{fig:A4}) pulse profiles appear to go back to a single-peak configuration. 
The pulses are very broad at this point and it is not clear if the emission is from one pole or two poles that are essentially blended together in the light curve. 
Later on, past MJD~58135 the accretion configuration changes again in that now the soft \textit{NICER} band stays single-peaked  (Fig.~\ref{fig:lcandprofiles} inset h) while the higher energy bands  return to a complex structure dominated by a single asymmetric peak. At the same time, past MJD~58135 GBM gets the double-peaked pulse profile structure back (Fig.~\ref{fig:A5}).

Such extreme luminosity-dependent character of the pulse profiles is usually interpreted in terms of a change in the accretion regime.
At lower luminosity the flow may be stopped primarily by coulomb collisions between particles as suggested by \cite{Basko+Sunyaev76} \citep[see also][]{Becker+12}.
In this regime, the flow stopping region is located near the neutron star surface where the plasma density is very high while the emission from the stopping region may be dominated by a pencil beam component.
However, once the accretion rate goes above a certain critical value, radiation pressure becomes the dominant stopping mechanism via photon scattering off of the electrons in the accretion flow.
This is when the radiation-dominated shock transition begins to work its way up the magnetic column, and an accretion column emerges, while a fan-beam emission pattern escaping the accretion column walls dominates the emission.

The critical luminosity $L_c$ that marks the transition between the coulomb dominated and radiation-pressure dominated accretion flow, that is the transition from the sub-critical to super-critical accretion regime, has been calculated by \citet{Becker+12} and \citet{Mushtukov+15}.
The computation of the exact value of critical luminosity strongly depends on a number of parameters that reflect the accretion physics and are currently not fully understood or constrained, such as the geometrical parameter $\Lambda$ characterizing whether the NS accretes from a wind
or from a disk, the shape of the photon spectrum inside the column, and the magnetic field strength $B$.
However, assuming canonical neutron star parameters, the standard critical luminosity is of the order of $L_c\sim10^{37}\,$erg\,s$^{-1}$.
This luminosity level is reached around MJD~58029 and 58166, very early and very late in the outburst, respectively (see Table~\ref{table:nicerobs}). In fact, the above described pulse profile changes show clear transitions around a much higher luminosity, $\sim 10^{38}\, $erg\, s$^{-1}$, around MJD~58037 and 58135, suggesting a much higher than typical value for the critical luminosity.
This finding is further corroborated by the spectral analysis and the study of the pulsed fraction discussed in the next sections, where additional observational evidence is gathered in support of the accretion regime transition scenario, and to further constrain the value $L_c$ of the critical luminosity.

\subsection{Hardness ratio evolution}\label{subsec:hardness_discussion}

Fig.~\ref{fig:soft_color} shows the hardness-intensity diagram for both the \textit{NICER} and GBM energy bands throughout the entire outburst of Swift~J0243.6+6124.
This diagram represents a model-independent way to probe the correlation between the spectral hardness and the observed source intensity, which has already been demonstrated for a number of accreting pulsars (e.g., \citealt{Reig+13}).
However, the exact mechanism that leads to the spectral hardness dependence on flux is not yet clear.
It is generally accepted that at the sub-critical regime ($L<L_c$), when the accretion rate grows the Coulomb layer that halts the accreting matter is pushed down into the accretion structure, where the electron gas becomes denser and hotter.
Within this scenario, radiation is scattered up to higher energies via the inverse Compton effect, and a harder spectrum emerges. Super-critical sources ($L>L_c$) show the opposite behavior, that is a spectral softening at higher flux levels.
This can be understood in the framework of two different scenarios, both of them involving the rise of an accretion column.
On the one hand, the turn-over to softer spectra at higher luminosity is due to the plasma temperature decrease upwards in the column \citep{Basko+Sunyaev76,Becker+12} which leads to the emission of softer photons.
On the other hand, the taller column leads to a lesser fraction of radiation reflected by the neutron star atmosphere (a process that induces spectral hardening), and therefore to a softer spectrum \citep{Postnov+15}.

Even if the exact mechanisms responsible for the spectral dependence on flux are not yet fully understood, \citet{Reig+13} showed that the turn-over present between the HB and the DB is happening at a luminosity level comparable to the critical luminosity.
Therefore, deriving the source luminosity at the turn-over level is of key importance. 

In the \textit{NICER} energy bands, the hardness ratio shows a hint of a turn-over near the beginning of the outburst, where a transition occurs from the HB to the DB (see Fig.~\ref{fig:soft_color}).
The farthest-right SC (soft color; Section~\ref{subsec:hardness} for \textit{NICER} occurred on MJD 58031.4.
Recognizing that \textit{NICER} observations are sparse and there is possible jitter in the SC value based on other sources in \citet{Reig+13}, we conservatively estimate that this turn-over occurred sometime between MJD 58031 and MJD 58037.
%, with a count rate of 78-310 cps in the 4-10 keV NICER band.
%As a simple conversion \citet{Jaisawal_2018} measured $L \sim 6.5 \times 10^{36}$ erg cm$^{-2}$ s$^{-1}$ in the 3-70 keV band with {\textit NuSTAR} on October 5, 2017.
%Scaling the NICER rates suggests this turn-over is occurring at $L \sim (0.7-2.6)\times 10^{37}$ erg cm$^{-2}$ s$^{-1}$ on the outburst rise. 
On the other hand, the red diamonds denoting the outburst decay in Fig.~\ref{fig:soft_color} show that if indeed we saw the turn-over during the outburst rise, we missed it on the decay. 
The more frequent observations obtained during the rise help to constrain the decay turn-over between MJD 58145 and MJD 58162.
%, with 4-10 keV count rates between 188 and 64 cps, respectively.
Comparing those events with our spectral results (see Sect.~\ref{subsec:spectral} and Table~\ref{table:nicerobs}), we derive a luminosity for the turn-over in the range  $ L_{0.1-10\,{\rm keV}}\sim (0.2-1.1) \times 10^{38}$ ergs s$^{-1}$.
We note that \citet{Jaisawal_2018} measured $ L_{3-70\,{\rm keV}}\sim 5.1\times 10^{37}$ erg s$^{-1}$ (corrected to d=7 kpc) with \textit{NuSTAR} on MJD 58031, thus \textit{NICER} and \textit{NuSTAR} derived luminosities are consistent within a factor of $\sim 2$. 
As a comparison, for typical values of the neutron star parameters, \citet{Becker+12} calculate the critical luminosity as
\begin{equation}
L_{\rm c}\sim1.5\times10^{37}\,B_{12}^{16/15}\, {\rm erg}\,{\rm s}^{-1} 
\label{eqn:lcrit}
\end{equation}
where $B_{12}$ is the magnetic field strength in units of $10^{12}\,$G, which is consistent with the value obtained in this work with \textit{NICER}, if the magnetic field is higher than the typical $\sim 10^{12}$G.

On the other hand, Fig.~\ref{fig:soft_color} shows that in the \textit{Fermi} GBM energy bands, the turn-over takes place at $\sim2\,$photons cm$^{-2}$ s$^{-1}$ keV$^{-1}$, corresponding to MJD $58040-58043$ and $58098-58105$, for the rising and the decay stages of the outburst, respectively.
GBM pulsed fluxes at the turn-over can be converted into a critical luminosity value by a simple scaling of the \textit{NuSTAR} observed luminosity.
In addition, as a consistency check, we also converted GBM pulsed fluxes into luminosity by comparing them with the \textit{Swift} BAT count rates in the same time windows, and using a conversion factor of $1.54\times10^{-7}\,$erg cm$^{-2}$\,cnt$^{-1}$ \citep{Doroshenko_2017}.
In the former case, the critical luminosity that results is  $ L_{3-70\, {\rm keV}} \sim (2.6-3.4)\times10^{38}\,$erg s$^{-1}$ for the rising stage of the outburst, and  $ L_{3-70\, \rm{keV}} \sim (2.2-3.4)\times10^{38}\,$erg s$^{-1}$ for the decay.
In the latter case, the resulting critical luminosity is $ L_{15-50\,{\rm keV}} \sim (3.2-4.3)\times10^{38}\,$erg s$^{-1}$ for the rising stage of the outburst, and  $ L_{15-50\, \rm{keV}} \sim (3.8-4.8)\times10^{38}\,$erg s$^{-1}$ for the decay.
In any case, the turn-over in the GBM energy bands happens at significantly higher luminosities than \textit{NICER}'s.
To the best of our knowledge, we excluded any instrumental origin for such an effect, thus concluding that the dependence of the turn-over luminosity on the energy band represents a physical effect.
An energy- and luminosity-dependent hardness ratio could, in fact, reflect a dependence of the spectral cutoff energy on the luminosity.
Previous works showed that, when the spectra of accreting pulsars are fit with a cutoff power-law model, the cutoff energy is directly correlated with the power-law photon index and, at the same time, inversely/directly correlated with the luminosity for the sub-/super-critical accretion regimes, respectively \citep{Ferrigno+13,Mueller+13,Reig+13,Malacaria+15}.
Since the cutoff energy is indicative of the electron temperature, its variability can be interpreted as the result of a Compton cooling efficiency dependence on the luminosity and the accretion regime.
Furthermore, we note that the typical cutoff energy in accreting X-ray pulsars is $E_{cut}\sim15-20\,$keV, that is around the GBM bands used to constrain the turn-over.
Following this qualitative scenario, the combination of luminosity-dependent cutoff energy and photon index can lead to a difference in the turn-over luminosity measured in different energy-bands.
However, we note that the energy bands used here to measure the turn-over can be affected by other factors, such as the Fe K line complex around $6-7\,$keV and the so-called ``10 keV feature'' \citep{Coburn+02}, commonly observed among accreting X-ray pulsars.

Considering these findings, we estimate a conservative value for the critical luminosity bracketed by the values separately derived with \textit{NICER} and GBM data, that is $(0.2-4.8)\times10^{38}\,$erg s$^{-1}$.
Given these bounds, the turn-over luminosity can be used to constrain the surface magnetic field value.
Following the approximate equation for critical luminosity provided by \citet{Becker+12} and cited above, 
%, that is $L_c\sim1.5\times10^{37}\,B_{12}^{16/15}\,$erg\,s$^{-1}$, 
we constrain the magnetic field to lie within the range $ (0.2-2.6)\times10^{13}\,$G.

\subsection{Constraints on the magnetic field from the QPO-like feature}

In the power spectra, a QPO-like feature at 50--70 mHz is found in the 0.2--12 keV band with \textit{NICER} early (MJD 58029--58031) and late (MJD 58166--58176) in the outburst. This feature can be used to estimate the magnetic field if we assume that the QPO frequency $\nu_{\rm QPO}$ is equal to the Keplerian orbital frequency $\nu_{\rm K}$ at distance $r_{\rm K}$, where
\begin{equation}
r_{\rm K} = \left[ \frac{GM}{(2 \pi \nu_{\rm K})^2}\right]^{1/3}.
\end{equation}
Next we assume that the magnetospheric radius is equal to the Alfven radius $r_{\rm A}$, where
\begin{equation}
r_{\rm A} = \left(\frac{\mu^4}{8G M \dot{M}^2}\right)^{1/7}
\end{equation}
and $\mu \approx BR^3$ is the magnetic dipole moment (see, e.g., \citealt{2002apa..book.....F}).
Setting $r_{\rm A} = r_{\rm K}$ yields
\begin{equation}
\mu = 4 \times 10^{31} {\rm G\ cm}^3 M_{1.4}^{5/6}\dot{M}_{-8}^{1/2} \left(\frac{\nu_{\rm K}}{50\,{\rm mHz}}\right)^{-7/6},
\end{equation}
where $M_{1.4} = 1.4\,M_{\odot}$ and $\dot{M}_{-8} = 10^{-8}\,M_{\odot}$ yr$^{-1}$. Finally we take $R = 10$~km to obtain
\begin{equation}
B = 4\times 10^{13}\,{\rm G}\ M_{1.4}^{-5/6} R_6^{-3} \dot{M}_{-8}^{1/2} \left(\frac{\nu_{\rm K}}{50\,\rm{mHz}}\right)^{-7/6}. 
\label{eqn:bfield}
\end{equation}
When the QPO was detected, the 0.1--10 keV luminosity was $(1.2-2.0) \times 10^{37} (d/7 \rm{kpc})^2$ for MJD 58029-58031. This corresponds to a mass accretion rate of $\dot{M} = (0.9-1.7) \times 10^{-9} M_{\odot}$ yr$^{-1}$. Substituting this rate and the QPO frequencies of 50 and 70 mHz respectively into Equation~\ref{eqn:bfield} yields $B\sim 1\times 10^{13} (d/7\,{\rm kpc})$ G. Substituting this value back into Equation~\ref{eqn:lcrit}, we obtain a critical luminosity of $L_{\rm c} \sim 1.7 \times 10^{38}$ erg s$^{-1}$ $(d/7\,{\rm kpc})^{16/15}$, consistent with and better constrained than that from the hardness ratio calculations.

\subsection{Pulsed fraction evolution}\label{subsec:pulse_fraction}

Another indication of changes happening at the times corresponding to the turn-over or, equivalently, to the critical luminosity, is found in the analysis of the pulsed fraction evolution (Fig.~\ref{fig:nicer_fraction}).
At the earliest, rising stage of the outburst, the pulsed fraction decreases with increasing flux, down to a local minimum around MJD $58037-58042$.
The pulsed fraction then rises up to a maximum around the outburst peak, then decreasing again during the decay stage, down to a second local minimum at MJD $58099-58106$, after which a new increase takes place, somehow symmetrically to the first half of the outburst. Above the critical luminosity, the pulse fraction clearly increases with increasing luminosity (See Figure~\ref{fig:nicer_fraction}, right panel), below the critical luminosity, especially as the outburst declines, the relationship is less clear. 
Once again, the local minimum occurrence times coincide with the turn-over times, and therefore with the critical luminosity, with the pulsed fraction changes closely tracking the pulse profile changes. 

The exact mechanism of the pulsed fraction dependence on the luminosity is not yet clear.
To complicate the matter further, the dependence is not unique: some sources exhibit pulsed fraction decreases with increasing luminosity (e.g., 4U~0115+63, \citealt{Tsygankov+07}; EXO~2030+375, \citealt{Epili+17}), while others show the opposite trend along with a non-monotonic evolution of the pulsed fraction with luminosity (see, e.g. V~0332+53, \citealt{Tsygankov+10}; SMC~X-2, \citealt{Jaisawal2016}).
Our observations suggest that the two opposing behaviors, that is the negative and positive dependence of the pulsed fraction on luminosity, are distinctive of sub- and super-critical accretion regimes, respectively.
However, we note that this is a simplistic scenario, and that more-detailed models need to be considered to take into account the diverse behavior of the pulsed fraction dependence on luminosity observed in different sources and at different accretion regimes.

\subsection{Swift J0243.6+6124 as the first known Galactic ULX Pulsar}

Ultraluminous X-ray sources (ULX) have been identified in a number of galaxies. A common definition is a luminosity $> 10^{39}$ erg s$^{-1}$ \cite{Kaaret2017}. At first, some of these systems were thought to contain intermediate mass black holes, $\sim 100-10000 M_{\odot}$ \citep{Sutton2012}. In 2014, the first ULX pulsar, M82 X--2, was discovered with \textit{NuSTAR} \citep{Bachetti2014}, with $L_{\rm 0.3-10 keV} = 1.8 \times 10^{40}$ erg s$^{-1}$ and a pulse period of 1.37 s. Since then, several other ULX pulsars have been discovered, including NGC 7793 P13, with a 0.42 s period and an observed peak luminosity of $\sim 10^{40}$ erg $s^{-1}$ \citep{Furst2016, Israel2017b}, and NGC 5907 ULX \citep{Israel2017a}, with a period of 1.13 s and a peak luminosity of $\sim 10^{41}$ erg s$^{-1}$. All three of these systems show rapid spin-up, have sinusoidal pulse profiles, and relatively low pulsed fractions that increase with increasing energy. They exhibit both bright and faint phases with luminosities of $\sim 10^{40-41}$ erg s$^{-1}$ and $\sim 10^{38}$ erg s$^{-1}$, respectively \citep{Kaaret2017}. All three systems were known as ULX sources before their identification as pulsars. Recently, a fourth ULX pulsar, NGC 300 ULX1, with an initial period of 32 s, rapidly spinning up, and a peak luminosity of $4.7 \times 10^{39}$ erg s$^{-1}$, was discovered \citep{Carpano2018}. This pulsar has a potential cyclotron feature at $\sim 13$ keV, suggesting a magnetic field of $B \sim 10^{12}$ G. \textit{Chandra} observations show a potential proton resonance scattering feature at 4.5 keV in M51 ULX8 that implies a neutron star surface magnetic field of $B \sim 10^{15}$ G \citep{Brightman2018}, although no periodicity has been reported in this case.

\begin{figure}
\plotone{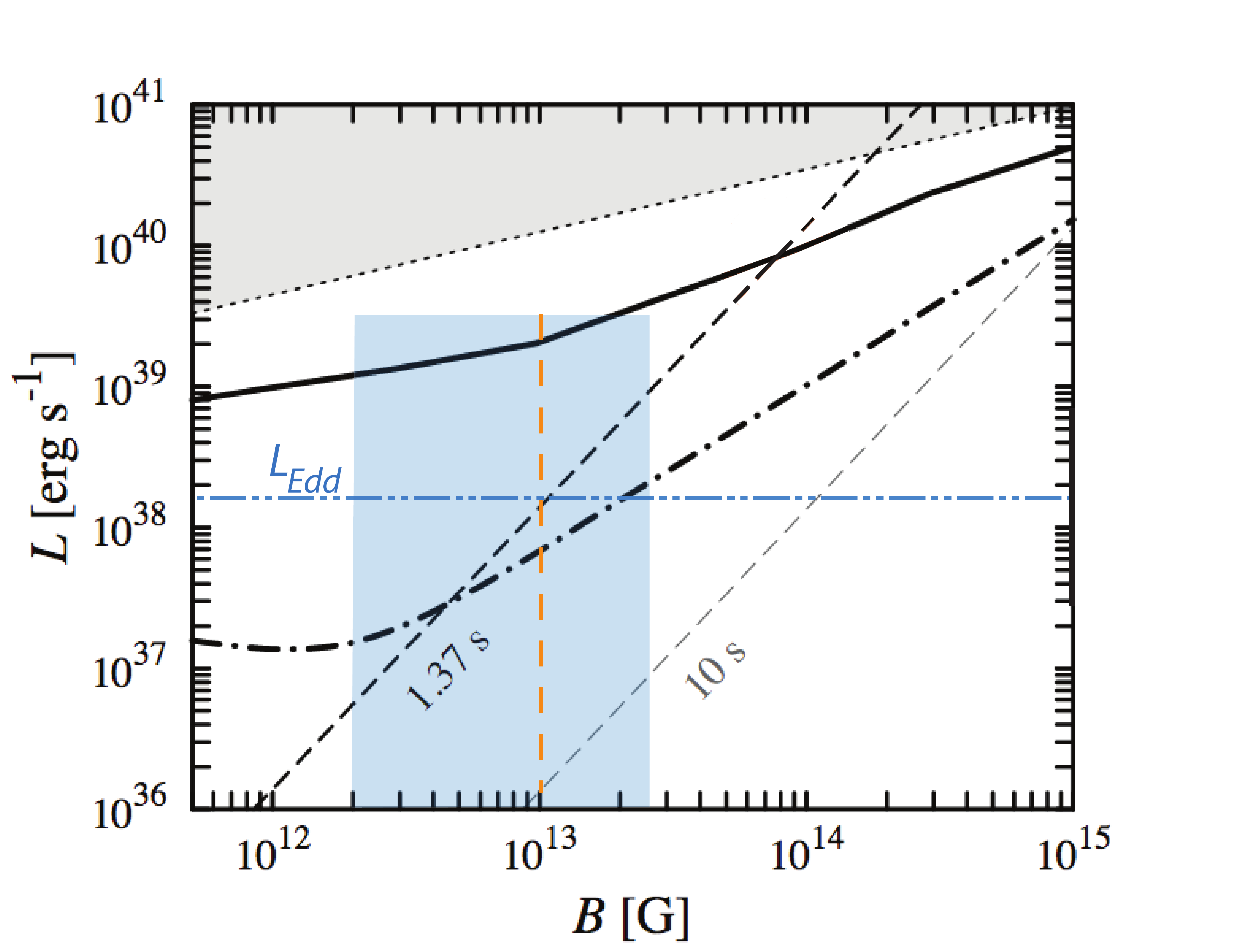}
\caption{Adapted from \citet{Mushtukov2015}. The solid black line represents the maximum luminosity for magnetized NSs as a function of magnetic field strength. The dotted line and shaded region above it indicates the upper limit where the spherization radius is smaller than the magnetospheric radius.  Dashed lines represent the lower limit on the X-ray luminosity due to the propeller effect. The dash-dotted line represents the critical luminosity function. The range of observed luminosity for Swift J0243 (for a distance of 7 kpc) is indicated by an orange vertical dashed-line at the magnetic field value derived from the QPO, and the blue-grey shaded box indicates the estimated range of magnetic field values derived from the critical luminosity (see text Sect. 5.2 and 5.3). For comparison, Eddington luminosity (for a 1.4 $M_{\odot}$ NS) is also indicated (long dash-double dotted horizontal turquoise line). \label{fig:ulx}}
\end{figure}

With a peak 0.1--10 keV luminosity of $2 \times 10^{39}$ erg s$^{-1}$, Swift J0243.6+6124 crosses the threshold for a ULX pulsar. Incorporating the \textit{Gaia} distance ranges, the dominant error on the luminosity, results in a peak luminosity of $(1.2-2.6) \times 10^{39}$ erg s$^{-1}$ for the \citet{Bailer-Jones2018} $\pm 1\sigma$ range of 5.7-8.4 kpc, and $(1.5-5.6) \times 10^{39}$ erg s$^{-1}$ for the 5-95\% confidence range of 6.3-12.3 kpc (M. Ramos, private communication), all above the $10^{39}$ erg s$^{-1}$ threshold for ULXs.  Swift J0243.6+6124 also shows rapid spin-up, exhibits an increasing pulsed fraction with energy, and has a relatively short spin period of $\sim 9.8$ s. \citet{Mushtukov2017} describe an optically thick envelope around accreting ULX pulsars that  produces a smooth pulse profile for $L>10^{39}$ erg s$^{-1}$ with a pulsed fraction that increases with energy. This optically thick envelope masks the observer's view of the neutron star surface. The profiles from Swift J0243.6+6124 are quite smooth at high luminosities but are more complex than the simple single peaked profiles observed from the other ULX pulsars, suggesting a different viewing geometry or that Swift J0243.6+6124 may be close to the threshold for this effect, similar to SMC X-3 \citep{Koliopanos+18}. \citet{Mushtukov2015} proposed that ULX sources and accreting pulsars are connected, with the brightest ULX pulsars having magnetar-like fields of $\sim 10^{14}$ G. They define an admissible luminosity range for accreting pulsars and ULX pulsars alike, where the maximum accretion luminosity can be approximated by $L_{\rm acc} \approx 0.35 B_{12}^{3/4} \times 10^{39}$ erg s$^{-1}$, for $10^{13}$ G $< B < 10^{15}$ G, shown in Figure~\ref{fig:ulx} as the solid dark line. They also define a minimum luminosity (denoted by a dashed line), based on the propeller effect which depends on the spin period of the pulsar. We demonstrate that Swift J0243.6+6124 is consistent with this picture, based on the luminosities measured with \textit{NICER} (keeping in mind that the total accretion luminosity is likely at least a factor of 2 higher to incorporate flux $>10$ keV) and  our estimate of Swift J0243.6+6124's magnetic field based on our critical luminosity estimate from the hardness ratios, pulse fraction, and pulse profiles, and  the magnetic field estimate from the QPO measured in with \textit{NICER}.

\section{Summary}\label{sec:summary}

%This paper reports on the observations of a particularly bright Be/X-ray transient source Swift J0243.6+6124 that was unknown as an X-ray binary until it suddenly erupted in late September 2017.
%The count rate observed with \textit{NICER} was unprecedented in that the source ultimately dwarfed for two months the Crab nebula and pulsar.
%The pulse profile analysis we have presented shows a complicated set of changes that are both luminosity and energy dependent. The power spectra are luminosity dependent as well, with a 50 mHz QPO at low luminosity with an evolving central frequency. 
Swift J0243.6+6124 underwent a giant outburst lasting about 150 days from October 2017 to February 2018, peaking well above Eddington luminosity at $ 1.8 \times 10^{39}$ erg s$^{-1}$ (0.1--10 keV).  This luminosity places it in the ULX category, making it the first known galactic ULX pulsar. Strong spin-up torques, high luminosities, and inferred mass accretion rates indicate that the pulsar was accreting from a disk, not a wind.

Near a luminosity of $10^{38}$ ergs s$^{-1}$ we identify a transition indicated by the following observations:
\begin{itemize}
\item{The pulse profiles (0.2--100 keV) evolve from single peaked to two distinct peaks in both \textit{NICER} and GBM data.}
\item{The pulsed fraction reaches a minimum and then increases with increasing intensity. Increases in pulsed fraction with energy are also seen, with the fraction approaching 100\% in the 8--12 keV band at the outburst peak.}
\item{The power spectra evolve with increasing intensity, with a QPO-like feature and a broad high frequency peak below $L \sim 10^{38}$ erg s$^{-1}$, becoming consistent with a single power-law in the brightest part of the outburst.}
\item{The hardness ratios become anticorrelated with intensity above $L \sim 10^{38}$ erg s$^{-1}$ in both \textit{NICER} and GBM data.}
\end{itemize}
All of these behaviors repeat as the source passes through the same luminosity ranges during the outburst decay. We interpret this as evidence for two accretion regimes that depend on
the accretion luminosity. The first regime shows a somewhat harder spectrum in the \textit{NICER} energy range and gives way to a somewhat softer spectrum as the source luminosity goes above $ \sim 10^{38}$ erg s$^{-1}$. We have identified the transition between these two regimes as the 
accretion structure on the neutron star surface transitioning from a Coulomb collisional stopping mechanism to a radiation-dominated stopping mechanism and have estimated that this occurs at a critical luminosity of $L_c = (0.2-4.8) \times 10^{38}$ erg s$^{-1}$, based on hardness ratios, and $L_c \sim 1.7 \times 10^{38}$ erg s$^{-1}$,  using the QPO features to estimate the magnetic field.
This lends support to models for the accretion flows put forward by \citet{Becker+12} and \citet{Mushtukov+15}. A critical luminosity of $~\sim 10^{38}$ erg s$^{-1}$ is the highest measured to date, suggesting a higher than average magnetic field of $\sim 10^{13}$ G for Swift J0243.6+6124.

\acknowledgements
This work was supported by NASA through the \textit{NICER} mission and the Astrophysics Explorers Program. This work was also supported by NASA through the Fermi Guest Investigator Program. GKJ acknowledges support from the Marie Sk{\l}odowska-Curie Actions grant no. 713683 (H2020; COFUNDPostdocDTU). 
This research has made use of data and software provided by the High Energy Astrophysics Science Archive Research Center (HEASARC), which is a service of the Astrophysics Science Division at NASA/GSFC and the High Energy Astrophysics Division of the Smithsonian Astrophysical Observatory.
Work at the Naval Research Laboratory by MTW, PSR, and MK was supported by NASA.
CM's research was supported by the NASA Postdoctoral Program at the MSFC, administered by USRA.

\bibliographystyle{aasjournal}
\bibliography{paper}

\facilities{NICER, Fermi, Swift}

\begin{appendices}
\section{Detailed Energy Dependent Pulse Profile Analysis}
\subsection{\textit{NICER} Pulse Profile Analysis}
Energy dependent \textit{NICER} pulse profiles were generated by selecting events in the 0.2--1, 1--2, 2--3, 3--5, 5--8, and 8--12 keV bands. Figures~\ref{fig:A1}--\ref{fig:A5} (top panels) show the evolution of the pulse profiles with time and with energy. Error bars are standard 1-$\sigma$ errors assuming Poisson statistics and are smaller than the line thickness in the brighter observations. Profiles are not phase connected. At lower average count rates ($<1000$ cps), early (Fig.~\ref{fig:A1}) and late  (Fig.~\ref{fig:A5}) in the outburst, considerable energy dependent pulse shape variations are observed and the pulse profile is very complex, with the largest peak closest to the deepest minimum.
Then a transition occurs at 500--1000 cps (Fig.~\ref{fig:A1} 4th--5th column, Fig.~\ref{fig:A5} 2nd--5th column) where the pulse profile becomes primarily a single asymmetric peak and seems to reverse in phase as to which peak is dominant. Above about 3000 cps (Figs.~\ref{fig:A1} last two columns \& \ref{fig:A3}), the profile becomes more symmetric and gradually splits into two equal peaks and the energy dependence becomes less dramatic. The minimum deepens between the two peaks as the intensity increases. As the outburst approaches its peak $>16000$ cps (Fig.~\ref{fig:A2}), the profile again becomes asymmetric with the peak closer to the deeper minimum again dominating. As the outburst decays, the profiles evolve very similarly through the shapes and complexities observed during the outburst rise. This analysis demonstrates the wealth of observations made possible by \textit{NICER} for an extremely bright source.

\subsection{GBM Pulse Profile Analysis}
Pulse profiles were generated in nine energy bands from 5--100 keV using GBM CTTE data. For each day of data, harmonic expansions including 24 sine and cosine terms in a pulse phase model were fitted to the GBM count rates. Times were barycentered and corrected for the orbital ephemeris described in the previous section. These harmonic expansions were converted to a more typical pulse profile and are plotted in Figure~\ref{fig:A1} to \ref{fig:A5}, bottom panels. Error bars are plotted for every $\sim 4$ bins, corresponding to the independent harmonics. Each profile has been shifted so that the minimum in the 8--12 keV band is at phase zero. Comparing these profiles to the \textit{NICER} profiles in the same columns, the higher energy profiles also evolve from complex at first to asymmetric and single peaked, but narrower, to double peaked, but more asymmetric than at lower energies, back to single  peaked, and complex. In the $0.2-1.0\,$keV band two distinct peaks emerge very quickly but in the higher energy range of $8-12\,$keV the emission is still dominated by one primary peak.
At even higher energy, in the GBM domain the pulse profile shows the distinct peaks already at early stages (See Figure~\ref{fig:A1}). When the profiles are the most complex, the phasing of the GBM and \textit{NICER} profiles may not be consistent.

% This will need to be updated if figures are added!
\renewcommand{\thefigure}{\figappendixprefix\arabic{figure}}
\setcounter{figure}{0}
\begingroup

\begin{figure}
\figurenum{A.1}
\vspace{-48pt}
\includegraphics[page=1,width=4.9in,angle=90]{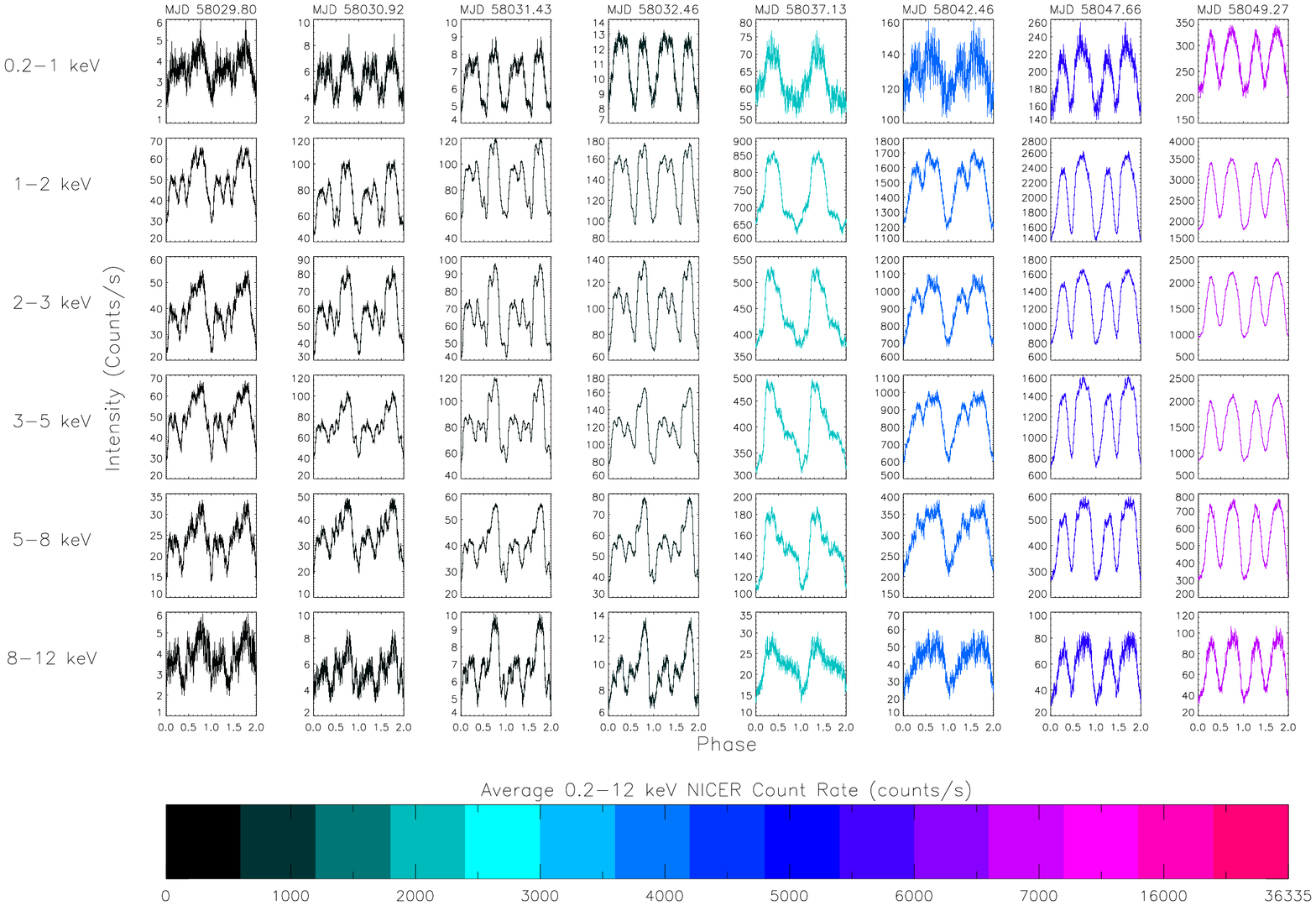}

\vspace{-60pt}

\includegraphics[page=1,width=4.9in,angle=90]{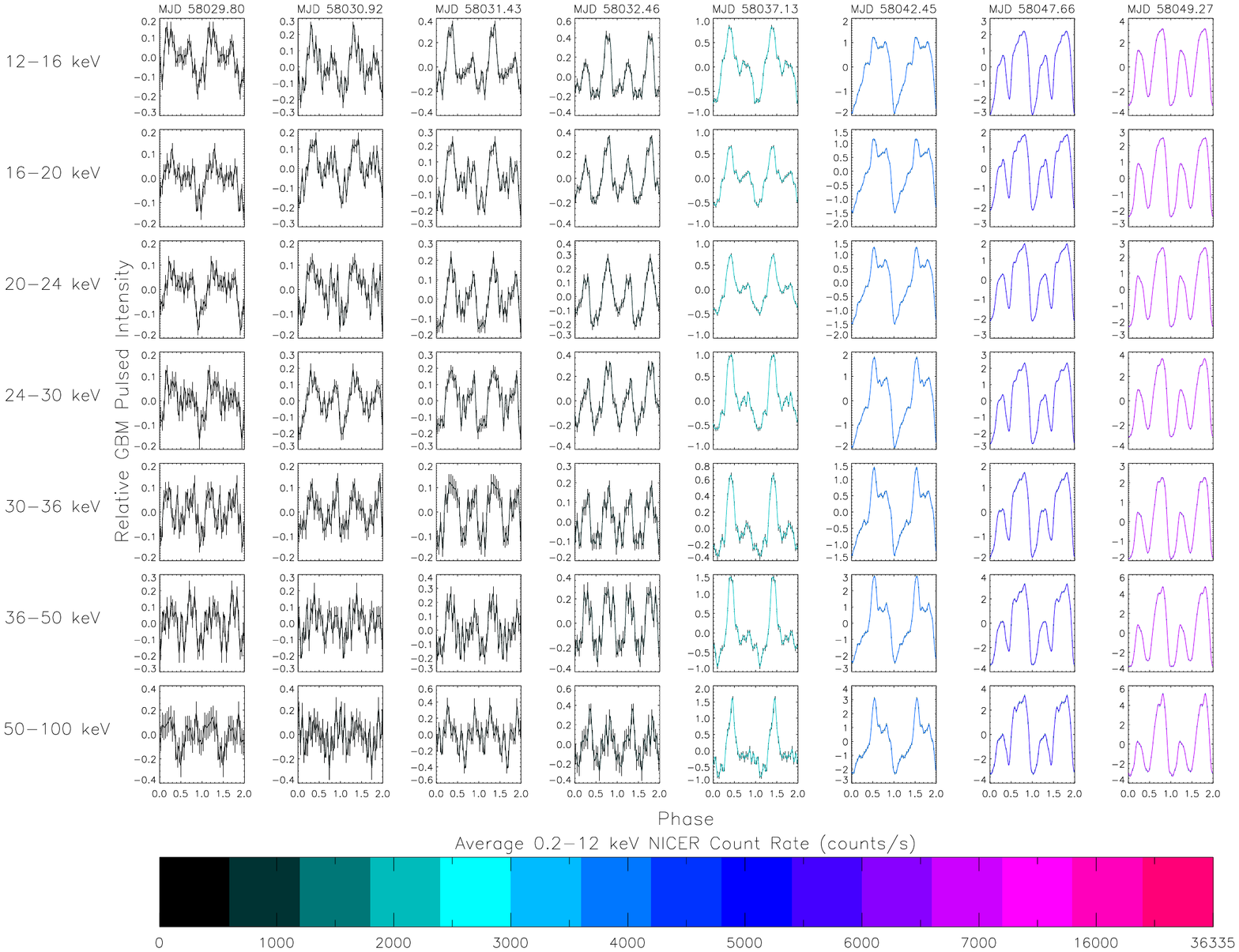}

\vspace{-12pt}
\caption{(Top panel): Swift J0243.6+6124 pulse profiles from \textit{NICER} observations in 6 energy bands. Profiles are arbitrarily aligned so that the minimum of the 0.2--12 keV profile is at phase 0.0. (Bottom panel) Swift J0243.6+6124 1-day pulse profiles measured with GBM in 7 energy bands. Profiles are arbitrarily shifted so that the minimum in the GBM 8--12 keV band (not shown) is at phase zero. (Both panels): The \textit{NICER} and GBM profiles are color coded using average \textit{NICER} 0.2--12 keV count rate for each observation. The \textit{NICER} and GBM profiles are folded with the same ephemeris, using the epoch in MJD that is the title of each column.  This figure includes observations from MJD 58029.8--58049.3, the rise of the outburst.
\label{fig:A1}}
\end{figure}

\begin{figure}
\figurenum{A2}
\vspace{-48pt}
\includegraphics[page=2,width=5.0in,angle=90]{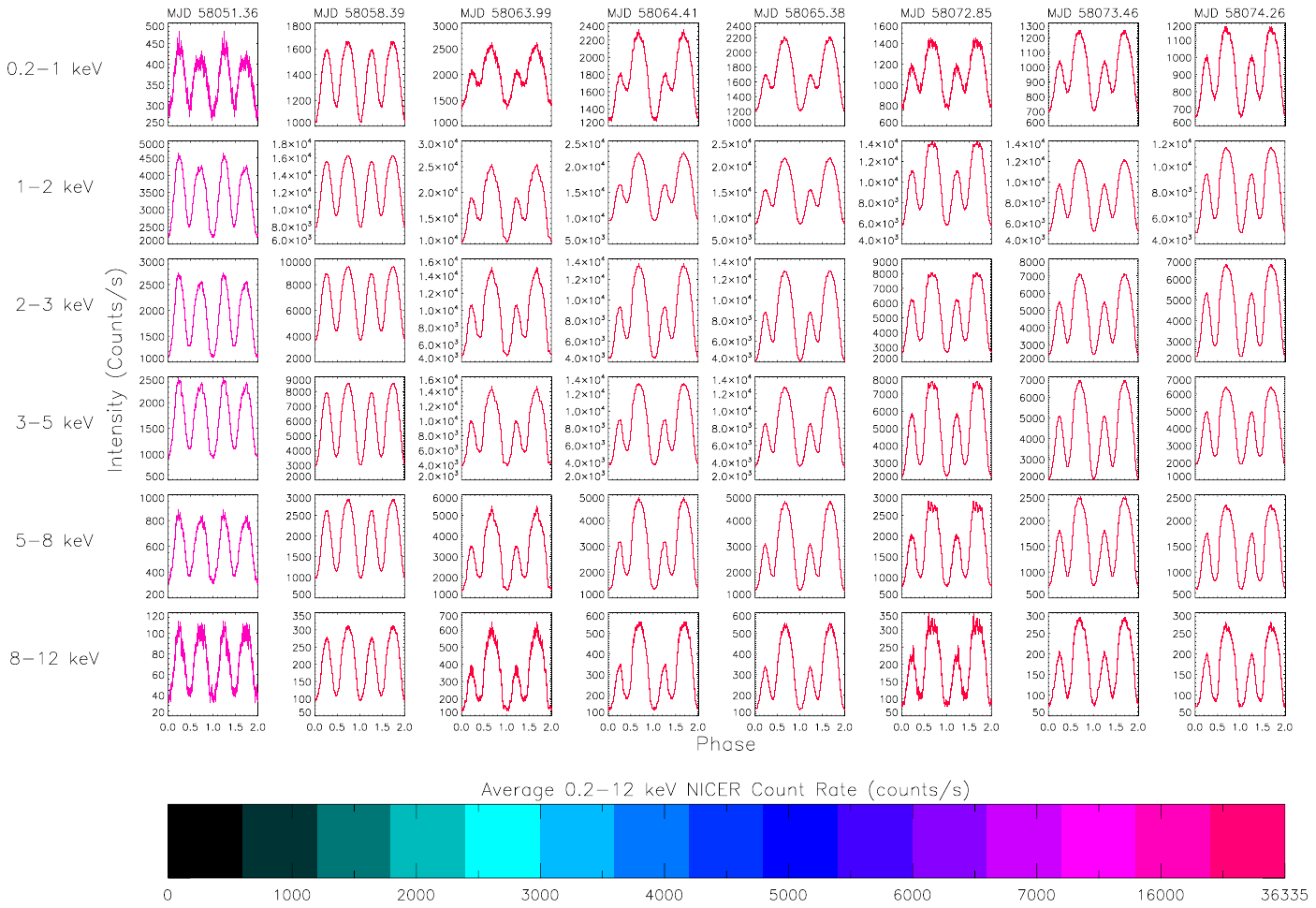}

\vspace{-60pt}

\includegraphics[page=2,width=5.0in,angle=90]{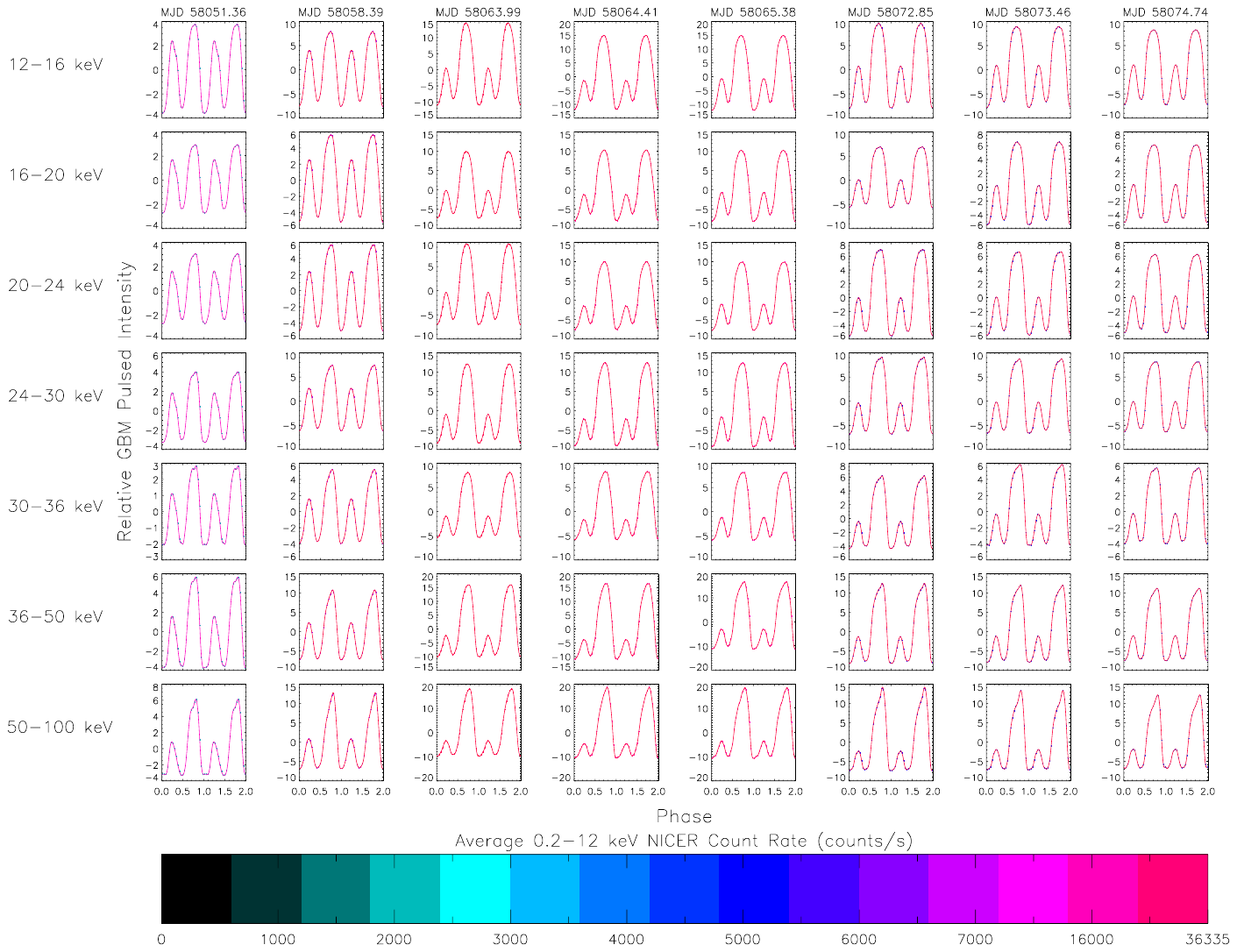}

\vspace{-12pt}
\caption{(Top panel): Swift J0243.6+6124 pulse profiles from \textit{NICER} observations in 6 energy bands. Profiles are arbitrarily aligned so that the minimum of the 0.2--12 keV profile is at phase 0.0. (Bottom panel) Swift J0243.6+6124 1-day pulse profiles measured with GBM in 7 energy bands. Profiles are arbitrarily shifted so that the minimum in the GBM 8--12 keV band (not shown) is at phase zero. (Both panels): The \textit{NICER} and GBM profiles are color coded using average \textit{NICER} 0.2--12 keV count rate for each observation. The \textit{NICER} and GBM profiles are folded with the same ephemeris, using the epoch in MJD that is the title of each column. The epoch in MJD is the title of each column. This figure spans MJD 58051.3-58074.7, including the peak of the outburst.
\label{fig:A2}}
\end{figure}

\begin{figure}
\figurenum{A.3}
\vspace{-48pt}
\includegraphics[page=3,width=5.0in,angle=90]{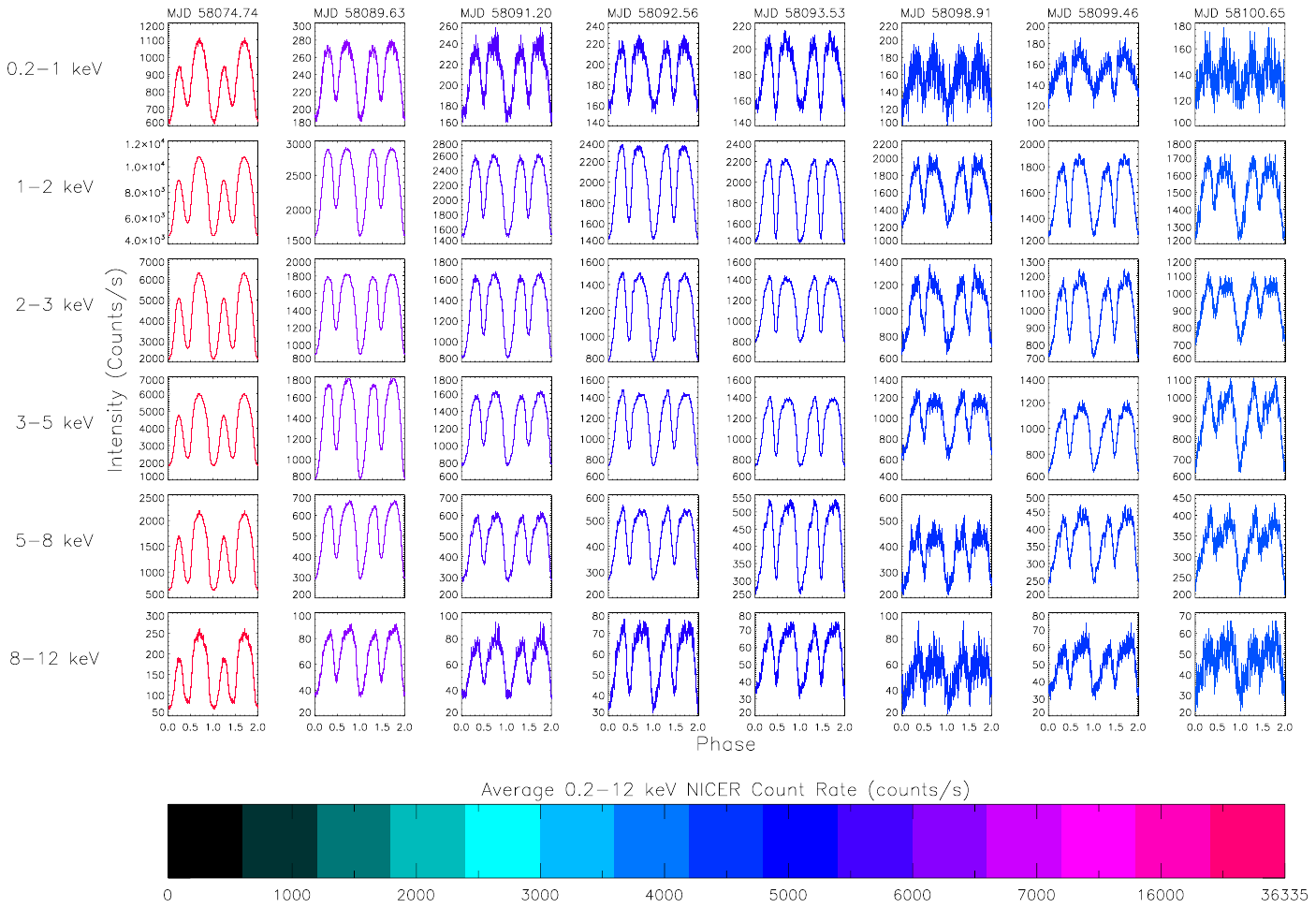}

\vspace{-60pt}

\includegraphics[page=3,width=5.0in,angle=90]{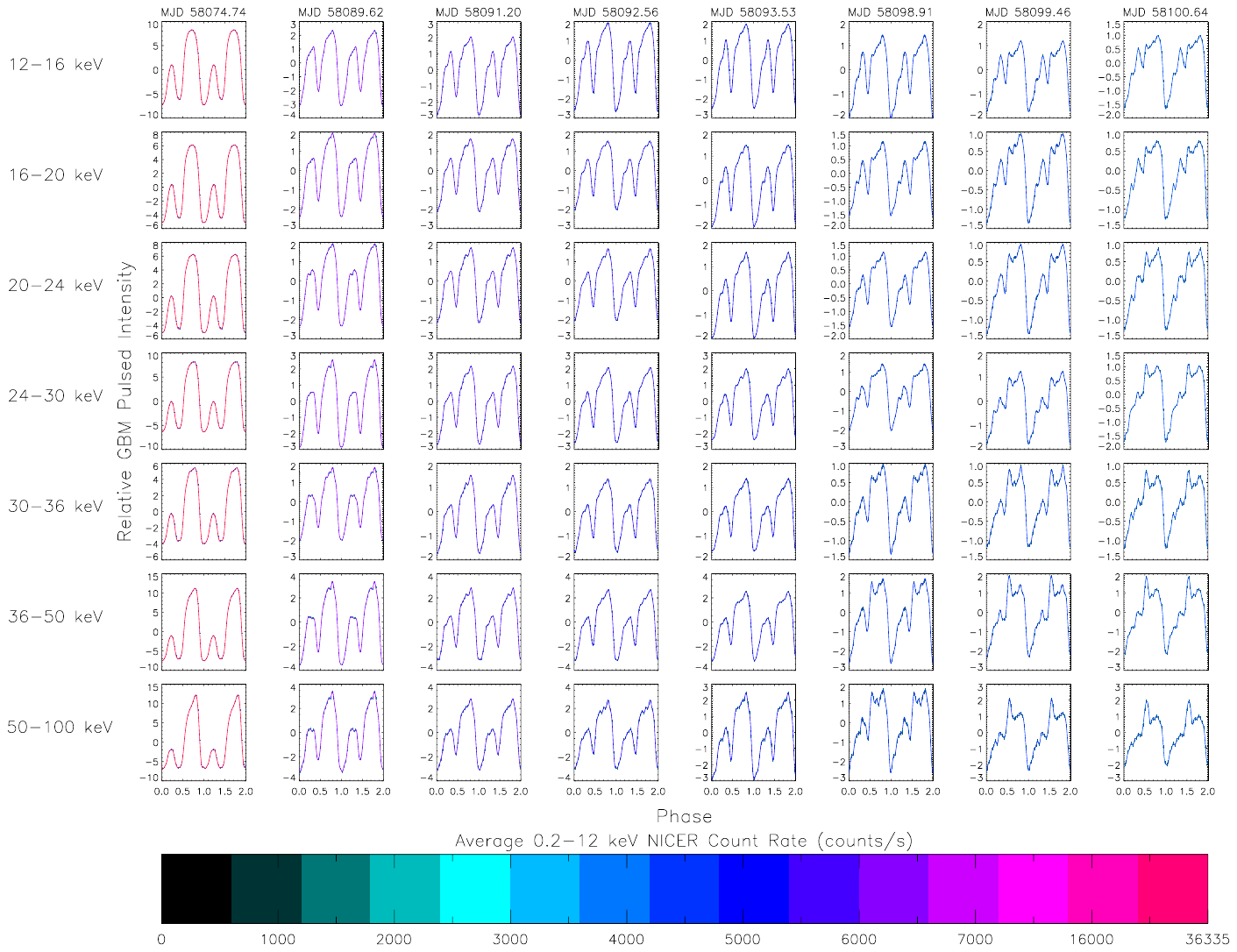}

\vspace{-12pt}
\caption{(Top panel): Swift J0243.6+6124 pulse profiles from \textit{NICER} observations in 6 energy bands. Profiles are arbitrarily aligned so that the minimum of the 0.2--12 keV profile is at phase 0.0. (Bottom panel) Swift J0243.6+6124 1-day pulse profiles measured with GBM in 7 energy bands. Profiles are arbitrarily shifted so that the minimum in the GBM 8--12 keV band (not shown) is at phase zero. (Both panels): The \textit{NICER} and GBM profiles are color coded using average \textit{NICER} 0.2--12 keV count rate for each observation. The \textit{NICER} and GBM profiles are folded with the same ephemeris, using the epoch in MJD that is the title of each column. This figure includes MJD 58074.7--58100.6, the beginning of the declining phase of the outburst.
\label{fig:A3}}
\end{figure}

\begin{figure}
\figurenum{A.4}
\vspace{-48pt}
\includegraphics[page=4,width=5.0in,angle=90]{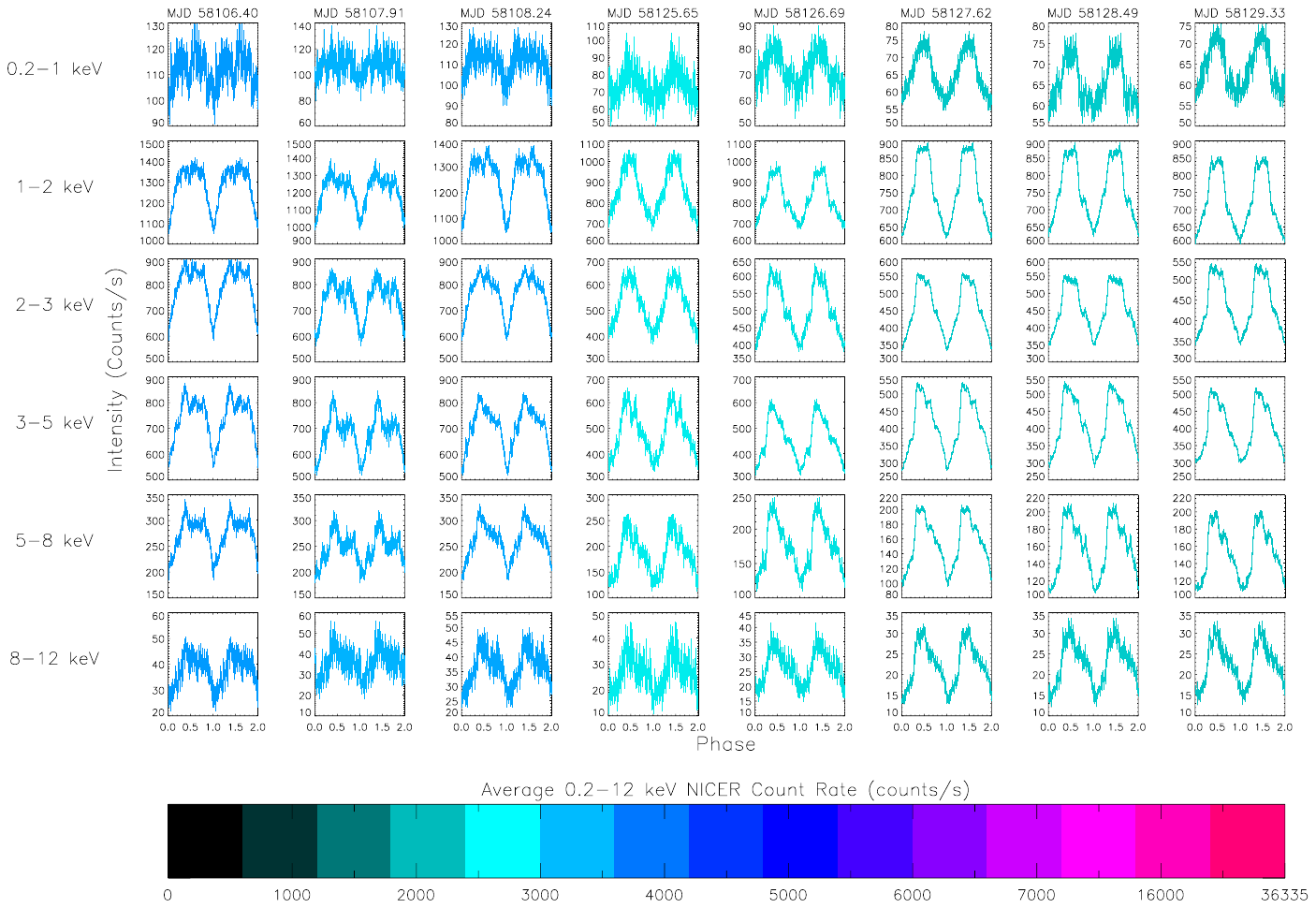}

\vspace{-60pt}

\includegraphics[page=4,width=5.0in,angle=90]{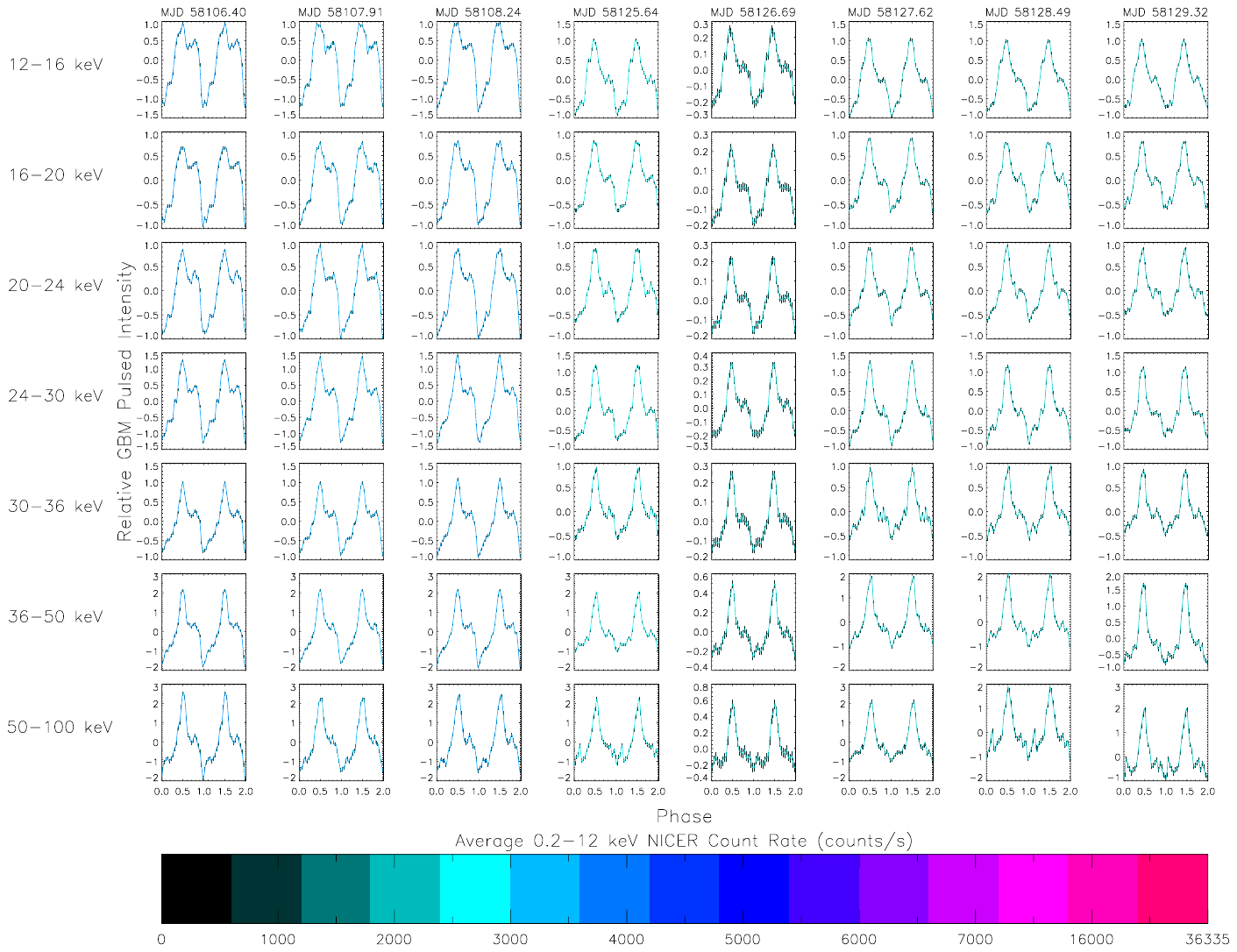}

\vspace{-12pt}

\caption{(Top panel): Swift J0243.6+6124 pulse profiles from \textit{NICER} observations in 6 energy bands. Profiles are arbitrarily aligned so that the minimum of the 0.2--12 keV profile is at phase 0.0. (Bottom panel) Swift J0243.6+6124 1-day pulse profiles measured with GBM in 7 energy bands. Profiles are arbitrarily shifted so that the minimum in the GBM 8--12 keV band (not shown) is at phase zero. (Both panels): The \textit{NICER} and GBM profiles are color coded using average \textit{NICER} 0.2--12 keV count rate for each observation. The \textit{NICER} and GBM profiles are folded with the same ephemeris, using the epoch in MJD that is the title of each column. This figure includes MJD 58106.4--58129.3, as the outburst continues to fade.
\label{fig:A4}}
\end{figure}

\begin{figure}
\figurenum{A.5}
\vspace{-54pt}

\includegraphics[page=5,width=5.0in,angle=90]{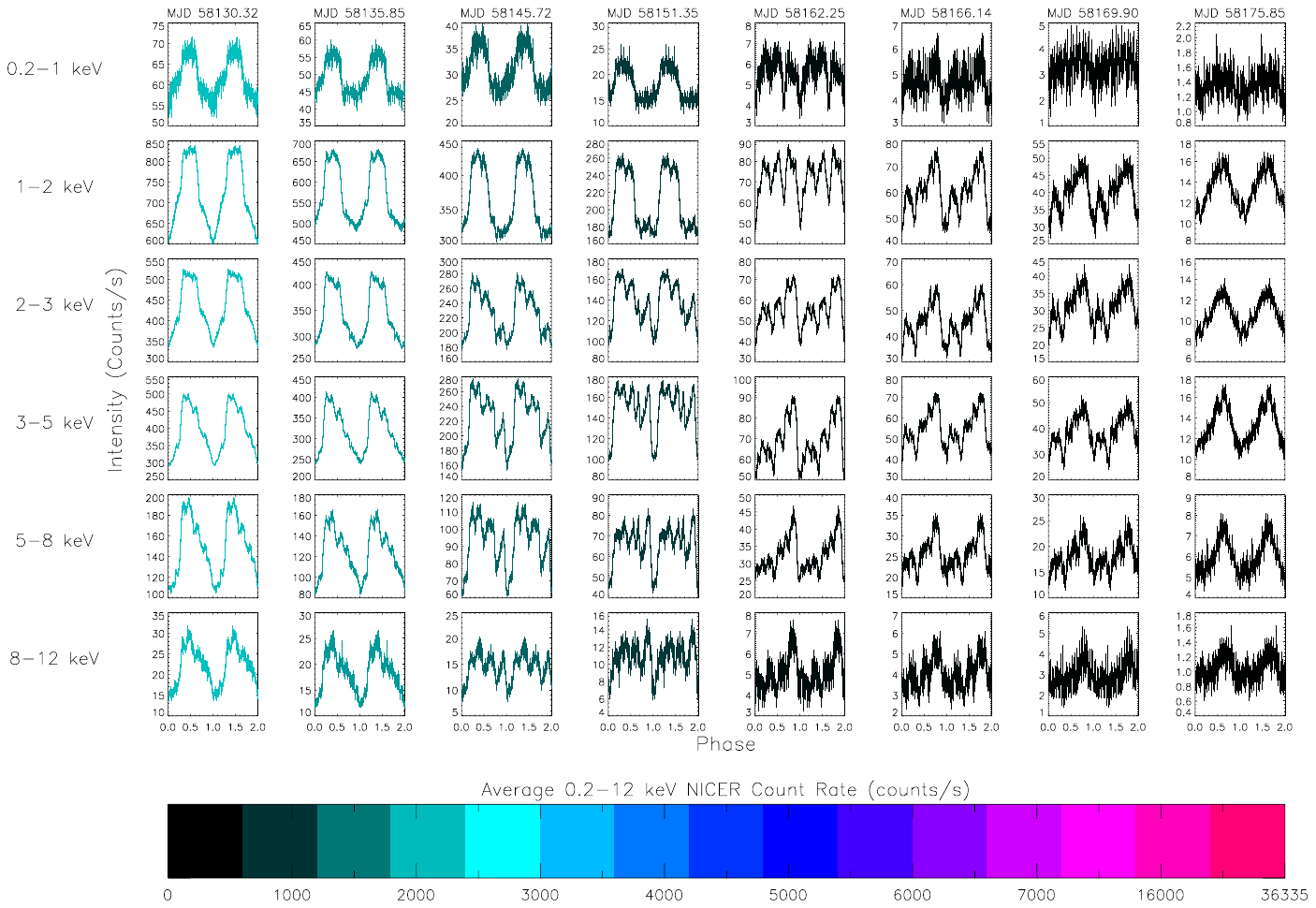}

\vspace{-60pt}

\includegraphics[page=5,width=5.0in,angle=90]{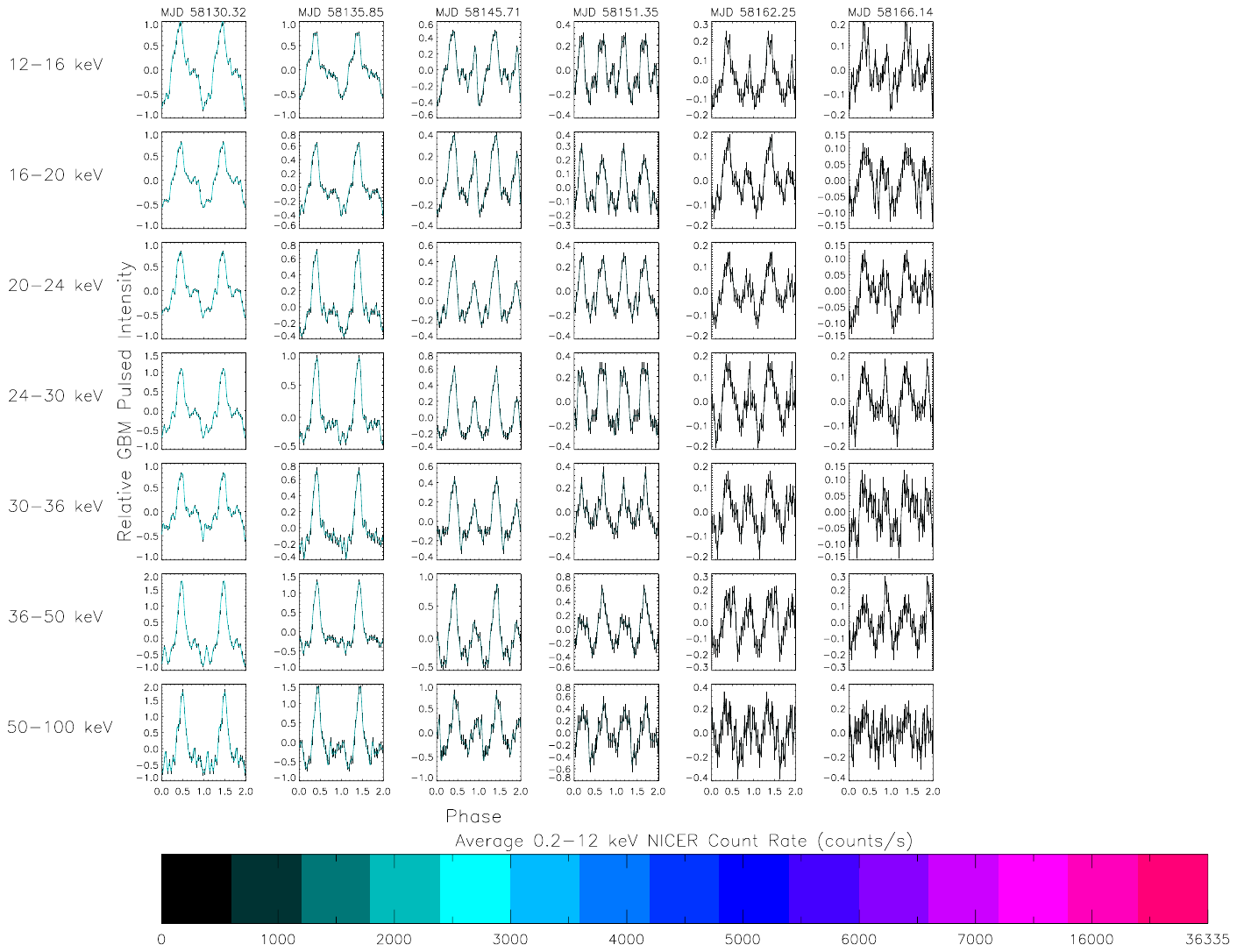}

\vspace{-24pt}

\caption{(Top panel): Swift J0243.6+6124 pulse profiles from \textit{NICER} observations in 6 energy bands. Profiles are arbitrarily aligned so that the minimum of the 0.2--12 keV profile is at phase 0.0. (Bottom panel) Swift J0243.6+6124 1-day pulse profiles measured with GBM in 7 energy bands. Profiles are arbitrarily shifted so that the minimum in the GBM 8--12 keV band (not shown) is at phase zero. (Both panels): The \textit{NICER} and GBM profiles are color coded using average \textit{NICER} 0.2--12 keV count rate for each observation. The \textit{NICER} and GBM profiles are folded with the same ephemeris, using the epoch in MJD that is the title of each column. This figure includes MJD 58130.3--58175.9 for \textit{NICER} and MJD 58130.3--58166.1 for GBM, as the outburst continues to slowly fade. Swift J0243.6+6124 dropped below GBM's one day detection threshold on MJD 58169.
\label{fig:A5}}
\end{figure}
\endgroup
\end{appendices}
\end{document}